\documentclass[preprint,aps,eqsecnum,amsmath,amssymb,floatfix]{revtex4}
\usepackage{graphicx}
\usepackage{psfrag}
\usepackage{subfigure}
\usepackage{tabls}
\usepackage[section]{placeins}
\newcommand{\be}{\begin{eqnarray}}
\newcommand{\ee}{\end{eqnarray}}
\newcommand{\ket}[1]{\vert\,{#1}\rangle}
\newcommand{\eps}{\epsilon}
\begin{document}
\begin{flushright}
SLAC-PUB-12096\\
UCRL-JRNL-340845
\end{flushright}
%\vspace*{1cm}
\title
{ Hadron Optics in
 Three-Dimensional Invariant Coordinate Space from Deeply Virtual Compton Scattering }

\author{\bf S. J. Brodsky$^a$, D. Chakrabarti$^b$, A. Harindranath$^c$, A.
Mukherjee$^d$, J. P. Vary$^{e,a,f}$}
%\email{dipankar@phys.ufl.edu}
\affiliation{$^a$ Stanford Linear Accelerator Center, Stanford
University, Stanford, California 94309, USA.\\
$^b$ Department of Physics, University of Florida, Gainesville,
FL-32611-8440, USA.\\
$^c$ Saha Institute of Nuclear Physics, 1/AF Bidhannagar, Kolkata
700064, India. \\
$^d$ Department of Physics,
Indian Institute of Technology, Powai, Mumbai 400076,
India.\\
$^e$ Department of Physics and Astronomy,
Iowa State University, Ames, Iowa 50011, USA.\\
$^f$ Lawrence Livermore National Laboratory, \\
L-414, 7000 East Avenue,
Livermore, California, 94551, USA.}

%\date{\today\\[2cm]}
\date{October 23, 2006}

\begin{abstract}

The Fourier transform of the deeply virtual Compton scattering
amplitude (DVCS) with respect to the skewness parameter  $\zeta= {
Q^2/ 2 p \cdot q}$ can be used to provide an image of the target
hadron in the boost-invariant variable $\sigma$, the coordinate
conjugate to light-front time $\tau={ t+ z/ c}$. As an illustration,
we construct a consistent covariant model of the DVCS amplitude and
its associated generalized parton distributions using the quantum
fluctuations  of a fermion state at one loop in QED, thus providing
a representation of the light-front wavefunctions  of a lepton in
$\sigma$ space. A consistent model for hadronic amplitudes can then
be obtained  by differentiating the light-front wavefunctions with
respect to the bound-state mass. The resulting DVCS helicity
amplitudes are evaluated as a function of $\sigma$ and the impact
parameter $\vec b_\perp$, thus providing a light-front image of the
target hadron in a frame-independent three-dimensional light-front
coordinate space.  Models for the LFWFs of hadrons in $(3+1)$
dimensions displaying confinement at large distances and conformal
symmetry at short distances have been obtained using the AdS/CFT
method. We also compute the LFWFs in this model in invariant three
dimensional coordinate space. We find that in the models studied,
the Fourier transform of the DVCS amplitudes exhibit diffraction
patterns. The results are analogous to the diffractive scattering of
a wave in optics where the distribution in $\sigma$ measures the
physical size of the scattering center in a one-dimensional system.
\end{abstract}
\maketitle

%%%%%%%%%%%%%%%%%%%%%%%%%%%%%%%%%%%%%%%%%%%%%%%%%%%%%%%%%%%%%%%%%%%%%%%%%%%
\section{Introduction}
%%%%%%%%%%%%%%%%%%%%%%%%%%%%%%%%%%%%%%%%%%%%%%%%%%%%%%%%%%%%%%%%%%%%%%%%%%%%%
Deeply virtual Compton scattering (DVCS), $\gamma^*(q)~ +~p(P) \to
\gamma(q')~+~ p(P^\prime),$ where the virtuality of the initial
photon $Q^2=-q^2$ is large, provides a valuable probe of the
elementary quark structure of the target proton near the light-cone.
At leading twist, QCD factorization applies \cite{fac}, and each
DVCS helicity amplitude factorizes as a convolution in $x$ of the
hard $\gamma^*q \rightarrow \gamma q$ Compton amplitude with a
hadronic sub-amplitude constructed from the Generalized Parton
Distributions (GPDs) $H(x,\zeta,t)$, $E(x,\zeta,t)$, $\tilde
H(x,\zeta,t)$ and $\tilde E(x,\zeta,t)$. Here $x$ is the light cone
momentum fraction of the struck quark; the skewness $\zeta={Q^2\over
2 P \cdot q}$ measures the longitudinal momentum transfer in the
DVCS process.

Measurements of the momentum and spin dependence of the  DVCS
process in $e^\pm ~ p \to \gamma ~e^\pm ~p$  can  provide a
remarkable window to the QCD structure of hadrons at the amplitude
level. The interference of the DVCS amplitude and the coherent
Bethe-Heitler amplitude leads to an $e^\pm$ asymmetry which is
related to the real part of the DVCS amplitude \cite{real}. The
imaginary part can also be accessed through various spin asymmetries
\cite{imag}. In the forward limit of zero momentum transfer, the
GPDs reduce to ordinary parton distributions; on the other hand, the
integration of GPDs over $x$ reduces them to electromagnetic and
gravitational form factors.

The DVCS helicity amplitudes can be constructed in light-cone gauge
from the overlap of the target hadron's light-front wavefunctions.
\cite{overlap1,overlap2}. Since the DVCS process involves
off-forward hadronic matrix elements of light-front bilocal
currents, the overlaps are in general non-diagonal in particle
number, unlike ordinary parton distributions. Thus in the case of
GPDs, one requires not only the diagonal parton number conserving $n
\to n$ overlap of the initial and final light-front wavefunctions,
but also an off-diagonal $n+1 \to n-1$ overlap, where the parton
number is decreased by two.  Thus the GPDs measure hadron structure
at the amplitude level in contrast to the probabilistic properties
of  parton distribution functions.

The GPDs have become objects of  much theoretical as well as
experimental attention since they provide a  rich source of
information of hadron structure. Burkardt has noted that a Fourier
Transform (FT) of the  GPDs with respect to the transverse momentum
transfer $\Delta_\perp$ in the idealized limit $\zeta=0$ measures
the impact parameter dependent parton distributions $q(x, b_\perp)$
defined from the absolute squares of the hadron's light-front wave
functions (LFWFs) in $x$ and
impact space \cite{bur1,bur2}. The impact representation on the
light-front was first introduced by Soper \cite{soper} in the
context of the FT of the elastic form factor (see Appendix
\ref{density}). The function $q(x, b_\perp)$ is defined for a hadron
with sharp plus momentum $P^+$, localized in the transverse plane,
such that the  transverse center of momentum vanishes $R_\perp=0$.
[ One can also work with a wave-packet  localized in the transverse
position space in order to avoid a state normalized to a
$\delta$-function. ] Thus $q(x, b_\perp)$ gives simultaneous
information on the distributions of a quark as a function of the
longitudinal light-front momentum fraction $x = k^+/P^+=
(k^0+k^3)/(P^0+P^3)$ and the transverse distance $b_\perp$ of the parton
from the center of the proton in the transverse plane. We use the
standard LF coordinates $P^\pm = P^0 \pm P^3,~ y^\pm = y^0 \pm y^3$.
Since the proton is on-shell, $P^+ P^- -P^2_\perp = M^2_p .$

Since the incoming photon is space-like ($q^2 < 0)$ and the final
photon is on-shell ($q'^2=0$), the skewness $\zeta$ is never zero in
a physical experiment.  In this paper, we will investigate the DVCS
amplitude in the longitudinal position space by taking the FT with
respect to $\zeta$.  We show that the FT of the DVCS amplitude in
$\zeta$ reveals the structure of a hadron target in a longitudinal
impact parameter space. Thus, our work is suited for the direct
analysis of experimental data and is complementary to the work of
Burkardt and Soper. Physically, the FT of the DVCS amplitude allows
one to measure the correlation within the hadron between the
incoming and outgoing quark currents at transverse separation
$b_\perp$ and longitudinal separation $\sigma = b^- P^+/2$ at fixed
light-front time $\tau = z+t/c.$ Since Lorentz boosts are
kinematical in the front form, the correlation determined in the
three-dimensional $b_\perp, \sigma$ space is frame-independent.

Even though light-front dynamics was proposed by Dirac
\cite{Dirac:1949cp} more than fifty years ago, and the utility of
the light-front momentum fraction $x = k^+ / P^+$  dates to the
inception of the Feynman parton model, very little is known about
the longitudinal coordinate space structure of hadron wave functions
and related physical observables. To the best of our knowledge, the
first work to investigate this subject is Ref. \cite{topo} where it
is shown that by Fourier transforming the form factors one observes
profiles  in $b^-$ with kinks and anti-kinks. In addition to the
light-front longitudinal structure of  DVCS amplitudes in one-loop
QED and meson models, we also present the corresponding structure of
the light-front wavefunctions (LFWFs) of the quantum fluctuations of
a lepton to order $e^2$ in QED.

Burkardt \cite{bur1} has noted the possibility of taking the FT with
respect to the longitudinal momentum of the active quark. However,
since the GPDs depend on a sharp $x$, the Heisenberg uncertainty
relation severely restricts the longitudinal position space
interpretation of GPDs. In contrast, we will deal directly with DVCS
amplitudes which are integrated over $x$ and take the FT with
respect to the longitudinal momentum {\em transfer}.

It has been shown in \cite{wigner} that one can define a quantum
mechanical Wigner distribution for the relativistic quarks and
gluons inside the proton. Integrating over $k^-$ and $k^\perp$, one
obtains a four dimensional quantum distribution which is a function
of ${\vec{r}}$ and $k^+$ where  ${\vec{r}}$ is the quark phase space
position defined in the rest frame of the proton. These
distributions are related to the FT of GPDs in the same frame.
However, the Wigner distributions cannot be measured experimentally.

In contrast, we will study the observable DVCS amplitudes directly
in longitudinal position space.  We shall show that the Fourier
transforms of the DVCS amplitudes in the variable $\sigma
=b^-P^+/2$, where the three-dimensional coordinate $\vec b=
(b_\perp, b^-)$ is conjugate to the momentum transfer $\vec \Delta$,
provides  a light-front image of the target hadron in a
frame-independent three-dimensional light-front coordinate space. We
find that in the models studied, the Fourier transform of the DVCS
amplitudes exhibit diffraction patterns. The results are analogous
to the diffractive scattering of a wave in optics where the
distribution in $\sigma$ measures the physical size of the
scattering center in a one-dimensional system.

A summary of our main results has been given in \cite{let}. In this
paper we will present a detailed analysis and provide several
additional results.

In order to illustrate the general framework, we will present an
explicit calculation of the FT of the DVCS on a fermion in QED at
one-loop order \cite{drell}. In effect, we shall represent a
spin-${1\over 2}$ system as a composite of a spin-${1\over 2}$
fermion and a spin-$1$ vector boson, with arbitrary masses
\cite{overlap1}. This one-loop model is self consistent since it has
the correct interrelation of different Fock components of the state
as given by the light-front eigenvalue equation \cite{rev}. In
particular, its two- and three-body Fock components can be obtained
analytically from QED. This model has been used to calculate the
spin and orbital angular momentum of a composite relativistic system
\cite{orbit} as well as the GPDs in the impact parameter space
\cite{dip1,dip2}.  The calculation is thus exact to $O(\alpha)$, and
it gives the Schwinger anomalous magnetic moment, the corresponding
electron's Dirac and Pauli form factors \cite{orbit,dip1} as well as
the correct gravitational form factors, including the vanishing of
the anomalous gravitomagnetic moment $B(0)$ in agreement with the
equivalence theorem \cite{grav}. In addition, it provides a template
for the wave functions of an effective quark-diquark model of the
valence Fock state of the proton light-front wave function.

Deep inelastic scattering structure functions and their connection
to the spin and orbital angular momentum of the nucleon have been
addressed for a dressed quark state in light-front QCD  in
\cite{oam,dis} using a similar Fock space expansion of the state.
This approach has also been used to investigate the twist-three GPDs
in \cite{marc}. We will also present here numerical results for a
simulated model for a meson-like hadron, which we obtain by taking a
derivative of the dressed electron LFWF with respect to the bound
state mass $M$, thus improving the behavior of the wave function
towards the end points in $x$. In this model, the DVCS amplitude is
purely real. A similar power-law LFWF has been used in
\cite{rad} to construct the GPDs for a meson.

In principle, the LFWFs  of hadrons in QCD can be computed using a
nonperturbative method such as Discretized Light Cone Quantization
(DLCQ) where the LF Hamiltonian is diagonalized on a free Fock
basis~\cite{rev}.  This has been accomplished for simple confining
quantum field theories such as $QCD(1+1) $~
\cite{Hornbostel:1988fb}.

Models for the LFWFs of hadrons in $(3+1)$ dimensions displaying
confinement at large distances and conformal symmetry at short
distances have been obtained using the AdS/CFT method~\cite{tera}.
We will also present the LFWFs in this hadron model in invariant
three dimensional coordinate space by Fourier transforming in both
$x$ and $k^\perp$.

The plan of the paper is as follows: Section II summarizes the
kinematics of the DVCS process.  In Section III we give the analytic
expressions for the DVCS amplitude and its explicit formulae in QED
at one loop. The calculation of the Fourier transform of the DVCS
amplitude for an electron target at one loop  is given in Section
IV. The simulated hadron model is discussed  in Section V. We then
derive the  DVCS amplitudes using a model meson LFWF as obtained
from holographic QCD  in Section VI. A summary and the conclusions
are given in Section VII.
%%%%%%%%%%%%%%%%%%%%%%%%%%%%%%%%%%%%%%%%%%%%%%%%%%%%%%%%%%%%%%%%%%%%%%%
\section{Kinematics}
%%%%%%%%%%%%%%%%%%%%%%%%%%%%%%%%%%%%%%%%%%%%%%%%%%%%%%%%%%%%%%%%%%%%%%%
The kinematics of the DVCS process has been given in detail in
\cite{overlap1,overlap2}. One can work in a frame where the momenta
of the initial and final proton has a  $\Delta \to -\Delta$ symmetry
\cite{overlap1}; however, in this frame, the kinematics in terms of
the parton momenta becomes more complicated. Here, we shall use the
frame of Ref. \cite{overlap1}.
The momenta of the initial and final proton are given by:
\begin{eqnarray}
P&=&
\left(\ P^+\ ,\ {\vec 0_\perp}\ ,\ {M^2\over P^+}\ \right)\ ,
\label{a1}\\
P'&=&
\left( (1-\zeta)P^+\ ,\ -{\vec \Delta_\perp}\ ,\ {M^2+{\vec
\Delta_\perp}^2 \over (1-\zeta)P^+}\right)\ ,
\end{eqnarray}
where $M$ is the proton mass.
%We use the component notation $V = (V^+,
%\vec{V}_\perp, V^-)$, and our metric is specified by $V^\pm = V^0 \pm
%V^z$ and $V^2 = V^+ V^- - {\vec V}_\perp^2$.
The four-momentum transfer from the target is
\begin{eqnarray}
\label{delta}
\Delta&=&P-P'\ =\
\left( \zeta P^+\ ,\ {\vec \Delta_\perp}\ ,\
{t+{\vec \Delta_\perp}^2 \over \zeta P^+}\right)\ ,
\end{eqnarray}
where $t = \Delta^2$. In addition, overall energy-momentum
conservation requires $\Delta^- = P^- - P'^-$, which connects ${\vec
\Delta_\perp}^2$, $\zeta$, and $t$ according to
\begin{equation}
 t \ = \ 2P\cdot \Delta\  =\
 -{\zeta^2M^2+{\vec \Delta_\perp}^2 \over 1-\zeta}\ .
 \label{na1anew}
\end{equation}
The coordinate $b$ conjugate to $\Delta$ is defined by $b\cdot
\Delta ={1\over 2}b^+\Delta^-+{1\over 2}b^-\Delta^+-b_\perp\cdot
\Delta_\perp$. We also define the boost invariant variable $\sigma =
b^- P^+/2 $  so that  ${1\over 2} b^-\Delta^+={1\over
2}b^-P^+\zeta=\sigma \zeta$. Thus $\sigma$ is an $`$impact parameter'
but in the boost-invariant longitudinal coordinate space.

It is convenient to choose
a frame where the incident space-like photon carries $q^+ = 0$ so that
$q^2= - Q^2 = - {\vec q_\perp}^{\;2}$ (however, it is not mandatory to
choose this frame):
\begin{eqnarray}
q&=& \left( 0\ ,\ {\vec q_\perp}\ ,\
{({\vec q_\perp}+{\vec \Delta_\perp})^2\over \zeta P^+}
+{\zeta M^2+{\vec \Delta_\perp}^2 \over (1-\zeta)P^+}\right)\ ,
\label{a2}
\\
q'&=&
\left( \zeta P^+\ ,\ {\vec q_\perp}+{\vec \Delta_\perp}\ ,\
{({\vec q_\perp}+{\vec \Delta_\perp})^2\over \zeta P^+}\right)\ .
\label{a2p}
\end{eqnarray}
We will be interested in deeply virtual Compton scattering, where
$Q^2$ is large compared to the masses and $-t$. Then, we have
\begin{equation}
{Q^2\over 2P \cdot q}=\zeta\
\label{nn3}
\end{equation}
up to corrections in $1/Q^2$. Thus $\zeta$ plays the role of the
Bjorken variable in deeply virtual Compton scattering. For a fixed
value of $-t$, the allowed range of $\zeta$ is given by
\begin{equation}
0\ \le\ \zeta\ \le\
{(-t)\over 2M^2}\ \ \left( {\sqrt{1+{4M^2\over (-t)}}}\ -\ 1\ \right)\ .
\label{nn4}
\end{equation}
%%%%%%%%%%%%%%%%%%%%%%%%%%%%%%%%%%%%%%%%%%%%%%%%%%%%%%%%%%%%%%%%%%%%%%%%%%
\section{Deeply Virtual Compton Scattering}
%%%%%%%%%%%%%%%%%%%%%%%%%%%%%%%%%%%%%%%%%%%%%%%%%%%%%%%%%%%%%%%%%%%%%%%%%%
%\subsection{Expression for the Deeply Virtual Compton Scattering Amplitude}
%%%%%%%%%%%%%%%%%%%%%%%%%%%%%%%%%%%%%%%%%%%%%%%%%%%%%%%%%%%%%%%%%%%%%%%%%%%%

The virtual Compton amplitude $M^{\mu\nu}({\vec q_\perp},{\vec
\Delta_\perp},\zeta)$, {\it i.e.}, the transition matrix element of
the process $\gamma^*(q) + p(P) \to \gamma(q') + p(P')$, can be
defined from the light-cone time-ordered product of currents
\begin{eqnarray}
&&M^{\mu\nu}({\vec q_\perp},{\vec \Delta_\perp},\zeta)\ =\
i\int d^4y\,
e^{-iq\cdot y}\langle P'|TJ^\mu (y)J^\nu (0)|P\rangle \ ,
\label{com1j}
\end{eqnarray}
where the Lorentz indices $\mu$ and $\nu$ denote the polarizations of
the initial and final photons respectively. In the limit
$Q^2\to \infty$ at fixed $\zeta$ and $t$ the Compton amplitude is thus
given by
\begin{eqnarray}
\lefteqn{
M^{IJ}({\vec q_\perp},{\vec \Delta_\perp},\zeta)\ =\
\epsilon^I_\mu\, \epsilon^{*J}_\nu\,
M^{\mu\nu}({\vec q_\perp},{\vec \Delta_\perp},\zeta)\ =\
- e^2_{q}\ {1 \over 2\bar P^+}
\int_{\zeta-1}^1{\rm d}z
}
\nonumber
\\
&\times& \left\{ \ {t}^{IJ}(z,\zeta)\ {\bar U}(P')
\left[
H(z,\zeta,t)\ {\gamma^+}
 +
E(z,\zeta,t)\ {i\over 2M}\, {\sigma^{+\alpha}}(-\Delta_\alpha)
\right] U(P) \right\} , \label{com1a}
\end{eqnarray}
where $\bar P={1\over 2}(P'+P)$. For simplicity we only consider one
quark with flavor $q$ and electric charge $e_{q}$. We here consider
the contribution of only the spin-independent GPDs $H$ and $E$.
Throughout our analysis we will assume the Born approximation to the
photon-quark amplitude; {\it i.e.}, the ``handbag" approximation,
corresponding to setting the Wilson line to $1$ in light-cone gauge.
In principle there can be rescattering corrections in the light-cone
gauge between the spectators at leading twist analogous to those
which occur in diffractive deep inelastic scattering
\cite{Brodsky:2002ue}, but these will not be considered here.

For circularly polarized initial and final photons ($I,\ J$ are
$\uparrow$ or $\downarrow)$) we have
\begin{eqnarray}
{t}^{\ \uparrow\uparrow}(z,\zeta)&=&
\phantom{-} \ {t}^{\ \downarrow\downarrow}(z,\zeta)\ =\
{1\over z-i\epsilon}\
+\ {1\over z-\zeta +i\epsilon}\ ,
\nonumber\\
{t}^{\ \uparrow\downarrow}(z,\zeta)
&=& {t}^{\ \downarrow\uparrow}(z,\zeta)
\ \ =\ \ 0\ \phantom{\frac{1}{2}}.
\label{com2p}
\end{eqnarray}
The two photon polarization vectors in light-cone gauge are given by
\begin{equation}
\epsilon^{\uparrow,\downarrow}=
\left(0\ ,\ \vec \epsilon^{\ \uparrow,\downarrow}_\perp\ ,\
{\vec \epsilon^{\ \uparrow,\downarrow}_\perp
 \cdot \vec k_\perp \over 2 k^+} \right)\ ,
\qquad
\vec \epsilon_\perp^{\ \uparrow,\downarrow}=
\mp {1\over\sqrt{2}} \left(
\begin{array}{c}
1 \\ \pm i
\end{array}
\right) \ ,
\label{p-vectors}
\end{equation}
where $k$ denotes the appropriate photon momentum.
The polarization
vectors satisfy the Lorentz condition $ k \cdot \epsilon =0$. For a
longitudinally polarized initial photon, the Compton amplitude is of
order $1/Q$ and thus vanishes in the limit $Q^2\to \infty$. At order
$1/Q$ there are several corrections to the simple structure in
Eq.~(\ref{com1a}). We do not consider them here.

The generalized parton distributions
 $H$, $E$ are defined through matrix elements
of the bilinear vector and axial vector currents on the light-cone:
\begin{eqnarray}
\lefteqn{
\int\frac{d y^-}{8\pi}\;e^{iz P^+y^-/2}\;
\langle P' | \bar\psi(0)\,\gamma^+\,\psi(y)\,|P\rangle
\Big|_{y^+=0, y_\perp=0}
} \hspace{2em} \nonumber \\
&=&
{1\over 2\bar P^+}\ {\bar U}(P') \left[ \,
H(z,\zeta,t)\ {\gamma^+}
 +
E(z,\zeta,t)\
{i\over 2M}\, {\sigma^{+\alpha}}(-\Delta_\alpha)
\right]  U(P)\ ,
\label{defhe}
\end{eqnarray}
The off-forward matrix elements given by Eq. (\ref{defhe}) can be
expressed in terms of overlaps of LFWFs of the state
\cite{overlap1,overlap2}. We now calculate the matrix elements in
terms of the LFWFs. For this, we take the state to be an electron in
QED at one loop and consider the LFWFs for this system.
%%%%%%%%%%%%%%%%%%%%%%%%%%%%%%%%%%%%%%%%%%%%%%%%%%%%%%%%%%%%%%%%%%%%%%%%%%%%
\subsection{DVCS in QED at one Loop}
%%%%%%%%%%%%%%%%%%%%%%%%%%%%%%%%%%%%%%%%%%%%%%%%%%%%%%%%%%%%%%%%%%%%%%%%%
The light-front Fock state wavefunctions corresponding to the
quantum fluctuations of a physical electron can be systematically
evaluated in QED perturbation theory. The light-cone time ordered
contribution for the state to the DVCS amplitude are given in Fig. 6
of \cite{overlap1}. The state is expanded in Fock space, giving
contributions from $\ket{e^- \gamma}$ and $\ket{e^- e^- e^+}$, in
addition to renormalizing the one-electron state. The two-particle
state is expanded as,
\begin{eqnarray}
\lefteqn{
\left|\Psi^{\uparrow}_{\rm two \ particle}(P^+, \vec P_\perp = \vec
0_\perp)\right> =
\int\frac{{\mathrm d} x \, {\mathrm d}^2
           {\vec k}_{\perp} }{\sqrt{x(1-x)}\, 16 \pi^3}
} \nonumber
\\
&&
\left[ \ \ \,
\psi^{\uparrow}_{+\frac{1}{2}\, +1}(x,{\vec k}_{\perp})\,
\left| +\frac{1}{2}\, +1\, ;\,\, xP^+\, ,\,\, {\vec k}_{\perp}\right>
+\psi^{\uparrow}_{+\frac{1}{2}\, -1}(x,{\vec k}_{\perp})\,
\left| +\frac{1}{2}\, -1\, ;\,\, xP^+\, ,\,\, {\vec k}_{\perp}\right>
\right.
\nonumber\\
&&\left. {}
+\psi^{\uparrow}_{-\frac{1}{2}\, +1} (x,{\vec k}_{\perp})\,
\left| -\frac{1}{2}\, +1\, ;\,\, xP^+\, ,\,\, {\vec k}_{\perp}\right>
+\psi^{\uparrow}_{-\frac{1}{2}\, -1} (x,{\vec k}_{\perp})\,
\left| -\frac{1}{2}\, -1\, ;\,\, xP^+\, ,\,\, {\vec k}_{\perp}\right>\
\right] \ ,
\label{vsn1}
\end{eqnarray}
where the two-particle states $|s_{\rm f}^z, s_{\rm b}^z; \ x, {\vec
k}_{\perp} \rangle$ are normalized as in \cite{overlap1}. Here
$s_{\rm f}^z$ and $s_{\rm b}^z$ denote the $z$-component of the
spins of the constituent fermion and boson, respectively, and the
variables $x$ and ${\vec k}_{\perp}$ refer to the momentum of the
fermion. The light cone momentum fraction $x_i= {k_i^+\over P^+}$
satisfy $0< x_i \le 1$, $\sum_i x_i =1$. We employ the light-cone
gauge $A^+=0$, so that the gauge boson polarizations are physical.
The three-particle state has a similar expansion. Both the two- and
three-particle Fock state components are given in \cite{overlap1}.
The two-particle wave function for spin-up electron are
\cite{orbit,drell,overlap1}
\begin{equation}
\left
\{ \begin{array}{l}
\psi^{\uparrow}_{+\frac{1}{2}\, +1} (x,{\vec k}_{\perp})=-{\sqrt{2}}
\ \frac{-k^1+{i} k^2}{x(1-x)}\,
\varphi \ ,\\
\psi^{\uparrow}_{+\frac{1}{2}\, -1} (x,{\vec k}_{\perp})=-{\sqrt{2}}
\ \frac{k^1+{i} k^2}{1-x }\,
\varphi \ ,\\
\psi^{\uparrow}_{-\frac{1}{2}\, +1} (x,{\vec k}_{\perp})=-{\sqrt{2}}
\ (M-{m\over x})\,
\varphi \ ,\\
\psi^{\uparrow}_{-\frac{1}{2}\, -1} (x,{\vec k}_{\perp})=0\ ,
\end{array}
\right \}~.
\label{vsn2}
\end{equation}
\begin{equation}
\varphi (x,{\vec k}_{\perp}) = \frac{e}{\sqrt{1-x}}\
\frac{1}{M^2-{{\vec k}_{\perp}^2+m^2 \over x}
-{{\vec k}_{\perp}^2+\lambda^2 \over 1-x}}\ .
\label{wfdenom}
\end{equation}
Similarly, the wave function for an electron with negative helicity
can also be obtained.

Following  references\cite{orbit,drell,overlap1}, we work in a
generalized form of QED by assigning a mass $M$ to the external
electrons, a distinct mass $m$ for the internal electron lines, and
a nonzero mass $\lambda$ for the internal photon lines, assuming the
stability condition $M<m+\lambda$. This provides a model for a
composite fermion state with mass $M$ with  fermion and  vector
``diquark" constituents. The electron in QED also has a one-particle
component
\begin{equation}
\left|\Psi_{\rm one \ particle}^{\uparrow , \downarrow}
(P^+, \vec P_\perp = \vec 0_\perp)\right> =
\int {{\mathrm d}x\, {\mathrm d}^2 {\vec{k}}_{\perp} \over
\sqrt{x}\, 16\pi^3}\ 16\pi^3 \delta (1-x)\,
{\delta}^2({\vec{k}}_{\perp})\
\psi_{(1)}\ \left| \pm {1\over 2} \, ;
xP^+, {\vec{k}}_{\perp} \right>
\label{bare1}
\end{equation}
where the single-constituent wavefunction is given by
\begin{equation}
\psi_{(1)} = \sqrt{Z} ~.
\label{oneparticle}
\end{equation}
Here $\sqrt Z$ is the wavefunction renormalization of the
one-particle state and ensures overall probability conservation.
Since we are working to ${\cal O} (\alpha) $, we can set $Z = 1$ in
the $3 \to 1$ wavefunction overlap contributions. At $x=1$, there
are contributions from the overlap of one particle states which
depend on $Z$. We have imposed  a cutoff on $x$ near this point.
Also, in order to regulate the ultraviolet
divergences, one has to introduce a regulator. Here, we use a cutoff
$\Lambda$ on the transverse momentum $k^\perp$ as a regulator.

In the domain $\zeta <z <1$, there are diagonal $2 \to 2$ overlap
contributions to Eq. (\ref{defhe}), both helicity flip,
$F^{22}_{+-}$ ($\lambda' \ne \lambda$) and helicity non-flip,
$F^{22}_{++}$ ($\lambda'=\lambda$) \cite{overlap1}. The GPDs
$H_{(2\to 2)}(z,\zeta,t)$ and $E_{(2\to 2)}(z,\zeta,t)$ are zero in
the domain $\zeta-1 < z < 0$, which corresponds to emission and
reabsorption of an $e^+$ from a physical electron. Contributions to
$H_{(n\to n)}(z,\zeta,t)$ and $E_{(n\to n)}(z,\zeta,t)$ in that
domain only appear beyond one-loop level since the DVCS amplitude
contains integrations over $z,~y^-,$ and $x$. When the integration
over $y^-$ is performed, the fermion part of the bilocal current
yields a factor $\delta(z-x)$ and the anti-fermion part of the
bilocal current yields a factor $\delta(z+x)$. The latter
contribution is absent in the one-loop DVCS amplitude of a electron
target, which we consider in the present work.

%%%%%%% xxxxxxx  %%%%%%%%%%%%%%%%%%%%%
%In the domain $\zeta <z <1$, there are diagonal $2 \to 2$ overlaps.
%These correspond to the helicity non-flip  and helicity flip
%contributions, respectively, given by \cite{overlap1},
We have,
\begin{eqnarray}
F^{22}_{++} &=&
{\sqrt{1-\zeta} \over 1-{\zeta\over 2}}\ H_{(2\to 2)}(x,\zeta,t)~
-~ {\zeta^2 \over 4 (1-{\zeta\over 2}){\sqrt{1-\zeta}}}
E_{(2\to 2)}(x,\zeta,t)
\nonumber  \\
&=&\int\frac{{\mathrm d}^2 {\vec k}_{\perp} }{16 \pi^3}
\Big[ \psi^{\uparrow *}_{+\frac{1}{2} +1}(x',{\vec k'}_{\perp})
\psi^{\uparrow}_{+\frac{1}{2} +1}(x,{\vec k}_{\perp})
+\psi^{\uparrow *}_{+\frac{1}{2} -1}(x',{\vec k'}_{\perp})
\psi^{\uparrow}_{+\frac{1}{2} -1}(x,{\vec k}_{\perp})
\nonumber\\
&&~~~~~~~~~~~~~~~~~~~~~~~~~~
+\psi^{\uparrow\ *}_{-\frac{1}{2} +1}(x',{\vec k'}_{\perp})
\psi^{\uparrow}_{-\frac{1}{2} +1}(x,{\vec k}_{\perp})
\Big]~ ,
\label{gf3} \\
\lefteqn{
F^{22}_{+-} =
{1 \over \sqrt{1-\zeta}}
{(\Delta^1-{i} \Delta^2)\over 2M} E_{(2\to 2)}(x,\zeta,t)
}\nonumber \\
&=&
\int\frac{{\mathrm d}^2 {\vec k}_{\perp} }{16 \pi^3}
\Big[\psi^{\uparrow *}_{+\frac{1}{2} -1}(x',{\vec k'}_{\perp})
\psi^{\downarrow}_{+\frac{1}{2} -1}(x,{\vec k}_{\perp})
+\psi^{\uparrow *}_{-\frac{1}{2} +1}(x',{\vec k'}_{\perp})
\psi^{\downarrow}_{-\frac{1}{2} +1}(x,{\vec k}_{\perp})
\Big]~ ,
\label{gf4}
\end{eqnarray}
where
\begin{equation}
x'={x-\zeta\over 1-\zeta},\ \ \
{\vec k'}_{\perp}={\vec k}^{~}_{\perp}-{1-x\over
1-\zeta}\ {\vec{\Delta}}_{\perp}\ .
\label{xprime}
\end{equation}
Using the explicit form of the two-particle wave functions, we obtain,
\be
F^{22}_{++} &=& {e^2\over 16 \pi^3} {1\over (1-x)} {[(1-\zeta)+
x (x-\zeta)]\over
\sqrt{1-\zeta}} \Big [I^{NF}_1+I^{NF}_2+[B(x,\zeta)+M^2 x (1-x)-m^2 (1-x)
\nonumber\\&&~~-\lambda^2 x] I^{NF}_3\Big ]
+{e^2\over 8 \pi^3} \Big ( {M\over 1-\zeta} -{m \over x-\zeta} \Big )
\Big ( M-{m\over x}\Big ) {x (1-x) (x-\zeta)\over \sqrt{1-\zeta}} I^{NF}_3.
\ee
We use the notation
\be
L_1 &=& (k^\perp)^2-2 k^\perp \cdot \Delta^\perp {(1-x)\over (1-\zeta)}
-B(x,\zeta),\nonumber\\
L_2  &=& (k^\perp)^2 -M^2 x (1-x) + m^2 (1-x) + \lambda^2 x
\ee
and $B(x,\zeta)= {M^2 (1-x) (x-\zeta)\over (1-\zeta)^2} -
{(\Delta^\perp)^2(1-x)^2\over (1-\zeta)^2} -m^2 {(1-x)\over (1-\zeta)}
-\lambda^2 {(x-\zeta)\over (1-\zeta)}.$
The integrals are given by
\be
I^{NF}_1 &=&\int {d^2k^\perp\over L_1}= \pi \log \Big [{\Lambda^2\over
\mid B(x,\zeta) \mid}\Big ]
\nonumber\\
I^{NF}_2 &=& \int {d^2k^\perp\over L_2}= \pi \log \Big [
{\Lambda^2\over \mid {-M^2 x (1-x) + m^2 (1-x) + \lambda^2 x}\mid}\Big ]
\nonumber\\
I^{NF}_3 &=& \int {d^2k^\perp\over L_1 L_2}= \pi \int_0^1 d \beta
{1\over D(x, \zeta, \beta)}
\ee
where $D(x,\zeta,\beta)=\beta m^2 (1-x) -\beta M^2 x (1-x)
+ \beta \lambda^2 x -(1-\beta) B(x,\zeta) -(1-\beta)^2 (1-x')^2
(\Delta^\perp)^2$.
Here $\Lambda$ is the cutoff on the transverse momentum $k^\perp$ and
$x'={(x-\zeta)\over (1-\zeta)} $.

The helicity-flip part can be written as,
\be F^{22}_{+-} &=&
{e^2\over 8 \pi^3} \Big [ {x (x-\zeta)\over \sqrt{1-\zeta}} \Big (
M-{m\over x}\Big ) I^F_1-\Big (M-{m\over x}\Big ) {x (1-x)
(x-\zeta)\over (1-\zeta)^{3/2}} I^F_2\Big ]
\nonumber\\&&~~-{e^2\over 8 \pi^3} \Big ( {M\over 1-\zeta}+{m\over\zeta-x}
\Big ) {x
(x-\zeta)\over \sqrt{1-\zeta}} I^F_1, \label{22f}
\ee
 where
$I^F_1=\int {d^2k^\perp k^\perp_V\over L_1 L_2}$ and $I^F_2=\int
{d^2k^\perp \Delta^\perp_V \over L_1 L_2}.$  We have used the
notation $ A^\perp_V=A^1-i A^2$. These integrals can be done using
the method described in Section III of Ref. \cite{dip2} and we
obtain
 \be I^F_1=\pi \int_0^1
dy~ {y~ (1-x') \Delta^\perp_V \over Q(x,\zeta,y)} \ee where
$Q(x,\zeta,y)= (1-y) [ -M^2 x (1-x) + m^2 (1-x) + \lambda^2 x] -y B
(x, \zeta) -y^2 (1-x')^2 (\Delta_\perp)^2$ and
 \be
I^F_2=\pi \int_0^1 {dy~ \Delta_V^\perp
\over Q(x,\zeta,y)}.
\ee
The scale $\Lambda$ dependence is suppressed in
$F^{22}_{+-}$.

The contribution in the domain, $0 < z < \zeta$ comes from an
overlap of three-particle and one-particle LFWFs. When the
electron's helicity is not flipped, this contribution is given by
\cite{overlap1},
\begin{eqnarray}
F^{31}_{++} &=&
{\sqrt{1-\zeta} \over 1- {\zeta\over 2}}\ H_{(3\to 1)}(x,\zeta,t)\,
-\, {\zeta^2 \over 4 (1-{\zeta\over 2}){\sqrt{1-\zeta}}}
E_{(3\to 1)}(x,\zeta,t)
\nonumber\\&=&
\sqrt{1-\zeta}\,
\int {{\rm d}^2{\vec{k}}_{\perp}\over 16\pi^3}\
\left[ \psi^{\uparrow}_{_{+{1\over 2}\,\,
+{1\over 2}\,\, -{1\over 2}}}
(x,1-\zeta,\zeta -x,\ {\vec k}_{\perp}, -{\vec \Delta}_{\perp},
{\vec \Delta}_{\perp}-{\vec k}_{\perp})
\right.
\nonumber\\
&& \hspace{6.4em} \left. {} +
\psi^{\uparrow}_{_{-{1\over 2}\,\, +{1\over 2}\,\, +{1\over 2}}}
(x,1-\zeta,\zeta -x,{\vec k}_{\perp},-{\vec \Delta}_{\perp},
{\vec \Delta}_{\perp}-{\vec k}_{\perp})
\right]\ ,\label{gf331}
\end{eqnarray}
and  can be
written as, using the three-particle wave function,
\be
F^{31}_{++}
&=& {e^2\over 8 \pi^3} (1-\zeta- \zeta x +x^2) {x~
\sqrt{1-\zeta}\over \zeta~ (1-x)} \Big [ {1\over (1-x)}~ J'^{NF}_1
-{1\over (1-\zeta)}~ J'^{NF}_2 \Big ] \nonumber\\&& -{e^2\over 8
\pi^3} {\sqrt{1-\zeta}~ x^2~ (\zeta-x)\over \zeta} \Big ( M-{m\over
x} \Big ) \Big ( {M\over (1-\zeta)} +{m \over (\zeta-x)} \Big )
J'^{NF}_3; \ee with \be J'^{NF}_1 &=& J^{NF}_2+[M^2 x (1-x) -m^2
(1-x) -\lambda^2 x ]
J^{NF}_3\nonumber\\
J'^{NF}_2 &=& {\zeta\over 2 x} \Big [ J^{NF}_2 - J^{NF}_1 + [M^2 x (1-x)
- m^2 (1-x) -\lambda^2 x + A(x, \zeta)] J^{NF}_3 \big ]\nonumber\\
J'^{NF}_3 &=& J^{NF}_3.
\ee
We denote
\be
l_1 &=& (k^\perp)^2-M^2 x (1-x) + m^2 (1-x) + \lambda^2 x,\nonumber\\
l_2 &=& (k^\perp)^2-2 k^\perp \cdot \Delta^\perp {x \over \zeta} + A(x,\zeta).
\ee
and $A(x, \zeta) = {1\over \zeta (1-\zeta)} \Big [ x (1-x) (\Delta^\perp)^2
+m^2 \zeta (1-\zeta)+ x~ \zeta~ (\zeta-x) M^2]$. The integrals can be
written as:

\be
J^{NF}_1 &=& \int {d^2 k^\perp\over l_1} = \pi \log \Big [
 {\Lambda^2 \over \mid  -M^2 x (1-x) + m^2 (1-x)
+\lambda^2 x \mid} \Big ] \nonumber\\
J^{NF}_2 &=&  \int {d^2 k^\perp\over l_2} = \pi \log
\Big [ {\Lambda^2 \over \mid A(x,\zeta)\mid}\Big ]\nonumber\\
J^{NF}_3 &=& \int {d^2 k^\perp\over l_1 l_2} = \pi
\int_0^1 d \beta {1\over C(x, \zeta, \beta)}.
\ee
where $C(x, \zeta, \beta)= -\beta M^2 x (1-x) + \beta m^2 (1-x)
+ \beta \lambda^2 x + (1-\beta) A(x,\zeta)-(1-\beta)^2
{x^2\over \zeta^2}\Delta_\perp^2$.

The helicity-flip part is given by \cite{overlap1},
\begin{eqnarray}
F^{31}_{+-}&=&{1 \over \sqrt{1-\zeta}}  \;
{(\Delta^1-{i} \Delta^2)\over 2M}\
E_{(3\to 1)}(x,\zeta,t)\nonumber\\
&=&
\sqrt{1-\zeta}\,
\int {{\rm d}^2{\vec{k}}_{\perp}\over 16\pi^3}\
\left[ \psi^{\downarrow}_{_{+{1\over 2}\,\,
+{1\over 2}\,\, -{1\over 2}}}
(x,1-\zeta,\zeta -x,\ {\vec k}_{\perp}, -{\vec \Delta}_{\perp},
{\vec \Delta}_{\perp}-{\vec k}_{\perp})
\right.
\nonumber
\\
&& \hspace{6.4em} \left. {} +
\psi^{\downarrow}_{_{-{1\over 2}\,\, +{1\over 2}\,\, +{1\over 2}}}
(x,1-\zeta,\zeta -x,{\vec k}_{\perp},-{\vec \Delta}_{\perp},
{\vec \Delta}_{\perp}-{\vec k}_{\perp})
\right]\ .\label{gf431}
\end{eqnarray}
Using the three-particle wave function, this can be written as,
\be
F^{31}_{+-} &=& {e^2\over 8 \pi^3} \sqrt{1-\zeta} \Big (M-{m\over x}
\Big ) {x^2 (\zeta-x)\over \zeta} \Big [ {1\over (1-\zeta)} J^F_2
-{1\over (1-x)} J^F_1 \Big ] \nonumber\\&&~~+ {e^2\over 8 \pi^3}
\sqrt{1-\zeta} \Big [ {M\over (1-\zeta)} +{m\over (\zeta-x)} \Big ]
{x^2 (\zeta-x)\over \zeta} {1\over (1-x)} J^F_1, \label{31f} \ee
where \be J^F_1  &=& \int {d^2 k^\perp k^\perp_V\over l_1 l_2}~ = ~
\pi \int_0^1 dy
{(1-y) \Delta^\perp_V {x\over \zeta}\over P(x, \zeta, y)}\nonumber\\
J^F_2 &=& \int {d^2 k^\perp \Delta^\perp_V\over l_1 l_2} ~=~ \pi \int_0^1 dy
{\Delta^\perp_V \over P(x, \zeta, y)},
\ee
with $P(x, \zeta, y)~=~ (1-y) A(x, \zeta) -(1-y)^2 {x^2\over \zeta^2}
(\Delta^\perp)^2 +y [-M^2 x (1-x) +m^2 (1-x) + \lambda^2 x]$.

We calculate the DVCS amplitude given by Eq. (\ref{com1a}) using the
off-forward matrix elements calculated above. The real and imaginary parts
are calculated separately using the prescription
\be
\int_0^1 dx {1\over x-\zeta+i \epsilon}F(x,\zeta)= P\int_0^1 dx
{1\over x-\zeta} F(x,\zeta) - i \pi F(\zeta,\zeta).
\ee
Here $P$ denotes the principal value defined as
\be
P \int_0^1 dx {1\over x-\zeta}F(x,\zeta)=\lim_{\epsilon \to 0}\Big[
\int_0^{\zeta-\epsilon}{1\over x-\zeta} F(x,\zeta)+\int_{\zeta+\epsilon}^1
{1\over x-\zeta} F(x,\zeta)\Big]
\ee
where
\be
F(x,\zeta)&=&F^{31}_{ij}(x,\zeta,\Delta_\perp),~{\rm for}~ 0<x<\zeta
\nonumber\\
         &=& F^{22}_{ij}(x,\zeta,\Delta_\perp),~{\rm for}~ \zeta<x<1\nonumber
\ee
 with $ij=++$ for helicity non-flip and $ij=+-$ for helicity flip amplitudes.
Since the off-forward matrix elements are continuous at $x=\zeta$,
  $F(\zeta,\zeta)=F^{22}_{ij}(x=\zeta,\zeta,\Delta_\perp)=
F^{31}_{ij}(x=\zeta,\zeta,\Delta_\perp)$.
Note that, for an electron state, the contribution vanishes for $x<0$ and
the principal value prescription cannot be used at $x=0$. The
off-forward matrix elements $F^{31}$ (which contribute in the kinematical
region $0 < x < \zeta$) vanish as $x \rightarrow 0$, as a result there is no
logarithmic divergence at this point for nonzero $\zeta$. But, we need to
be careful here as
when we consider the Fourier transform in $\sigma $ space, $\zeta$ can go
to zero and divergences from small $x$ can occur from $F^{22}$ which is
finite and nonzero
at $x,\zeta \to 0$.

The imaginary part of the amplitude when the electron helicity is
not flipped is then given by
\be {\mathrm{Im}} [M_{++}] (\zeta,
\Delta_\perp) =
   \pi e^2 F^{22}_{++}(x=\zeta,\zeta,\Delta_\perp).
\label{330}
\ee
A similar expression
holds in the case when the electron helicity is flipped $
({\mathrm{Im}} [M_{+-}] (\zeta, \Delta^\perp)) $ in which $ F_{++}$
are replaced by $F_{+-}$. The helicity-flip DVCS amplitude is
proportional to $(\Delta_1-i \Delta_2)$ as seen from  Eqs.
(\ref{22f}) and (\ref{31f}). In the numerical results for the helicity flip
processes that we  present here, for simplicity we
have taken $\Delta_2=0$.  The imaginary part receives
contributions
from $x=\zeta$. The off-forward matrix elements are continuous at $x=\zeta$,
and in Eq. (\ref{330}) we have used this continuity.
The other regions of $x$ contribute to the
real part. It is to be emphasized that we are using the handbag
approximation of the DVCS amplitude. Contributions from the Wilson
lines are in general not zero, and they can give rise to new phase
structures as seen in single-spin asymmetries~\cite{Brodsky:2002cx}.

The real  part of the DVCS amplitude is given by,
\be
{\mathrm{Re}}~
[M_{++}]~ (\zeta, \Delta_\perp) &=& -e^2  \int_\epsilon^{\zeta
-\epsilon_1} ~dx~
F^{31}_{++} (x,\zeta, \Delta_\perp) ~\Big [~  ~{1\over x}~
 +  ~{1\over{x-\zeta}}~  \Big ] \nonumber\\
&&- e^2
\int_{\zeta+\epsilon_1}^{1-\epsilon} ~dx~ F^{22}_{++} (x,\zeta,
\Delta_\perp) \Big [~{1\over x}~ + ~{1\over{x-\zeta}}~ \Big ]~.
\ee
Similar expression holds for the helicity flip DVCS amplitude.

 The energy dependence of the DVCS amplitude at the high energies  
$s >> Q^2, -t$ can be deduced up to logarithms using Regge  analysis. 
In  our  QED model  for DVCS, two spin $1/2$ propagators  are  
exchanged in the t-channel. This implies Regge behavior 
$s^{\alpha(0)}$ where $\alpha(0) = j_1 + j_2 -1 = 0.$    
  Thus, up to logarithms, the DVCS    amplitude has no dependence on
the initial energy of the incident electron and photon.

%%%%%%%%%%%%%%%%%%%%%%%%%%%%%%%%%%%%%%%%%%%%%%%%%%%%%%%%%%%%%%%%%%%
\section{Calculation of the  Fourier Transform}
\begin{figure}
\centering
\includegraphics[width=11cm,height=7cm,clip]{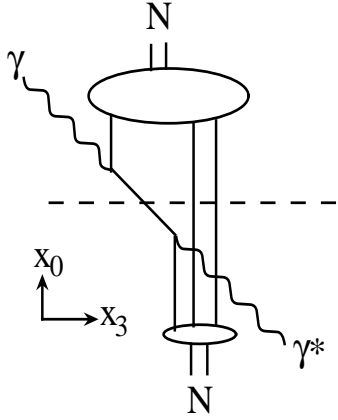}%
\caption{\label{fig1a} The handbag diagram for DVCS amplitude viewed
in coordinate space. The position of the struck quark differs by $x^-$
in the two wave functions (whereas $x^+=x^\perp=0$).}
\end{figure}

%%%%%%%%%%%%%%%%%%%%%%%%%%%%%%%%%%%%%%%%%%%%%%%%%%%%%%%%%%%%%%%%%%%
In order to obtain the DVCS amplitude in $b^-$ space, we take a
Fourier transform in $\zeta$ as,
\be
A_{++} (\sigma, t) =
{1\over 2 \pi } \int_{\epsilon_2}^{1-\epsilon_2} d\zeta e^{{i}
\sigma \zeta }~ M_{++}
(\zeta, \Delta_\perp),\nonumber\\
A_{+-} (\sigma, t) = {1\over 2 \pi }
\int_{\epsilon_2}^{1-\epsilon_2} d\zeta e^{{i} \sigma \zeta }~
M_{+-} (\zeta, \Delta_\perp),
\ee
where $\sigma = {1\over 2}P^+ b^-$ is the boost invariant longitudinal
distance on the light-cone. The spatial properties of deep inelastic
scattering obtained from a Fourier
transform of structure functions from $x= k^+/P^+$ to $b^-$ space has been
discussed by Hoyer \cite{Hoyer:2006xg}. In Fig. \ref{fig1a} we show the
handbag diagram of the DVCS amplitude in coordinate space, which is similar
to Fig. 10 of the above reference.

Both real and imaginary parts of the DVCS amplitude are  obtained
separately. The real part of the amplitude depends on the cutoffs.
Since the off-forward matrix elements are continuous at $x=\zeta$,
the DVCS
 amplitude is independent of the cutoff $\epsilon_1$.
The cutoffs have to be chosen such that $\eps_2-\eps_1 \ge \eps$,
$\eps_2+\eps_1 < 1-\eps$, in order to have the correct principal
value integration. In our numerical analysis, we have taken
$\eps=\eps_1=\eps_2/2=0.001$. As stated before, the cutoff at $x=0$ is
imposed for the numerical calculation and has a small effect on the result.
If, instead of imposing a cutoff
on transverse momentum, $\Lambda$, we imposed a cutoff on the invariant
mass \cite{orbit}, then the divergences at $x=1$ would have been
regulated by a non-zero photon mass \cite{zhang}.
The DVCS amplitude at $x=1$
 also receives a
contribution from the single particle sector of the Fock space
\cite{overlap1,marc,dip1,dip2}, which we did not take into account.
A detailed
 discussion about the cutoff scheme is provided in
Appendix \ref{cutoff}.

\begin{figure}
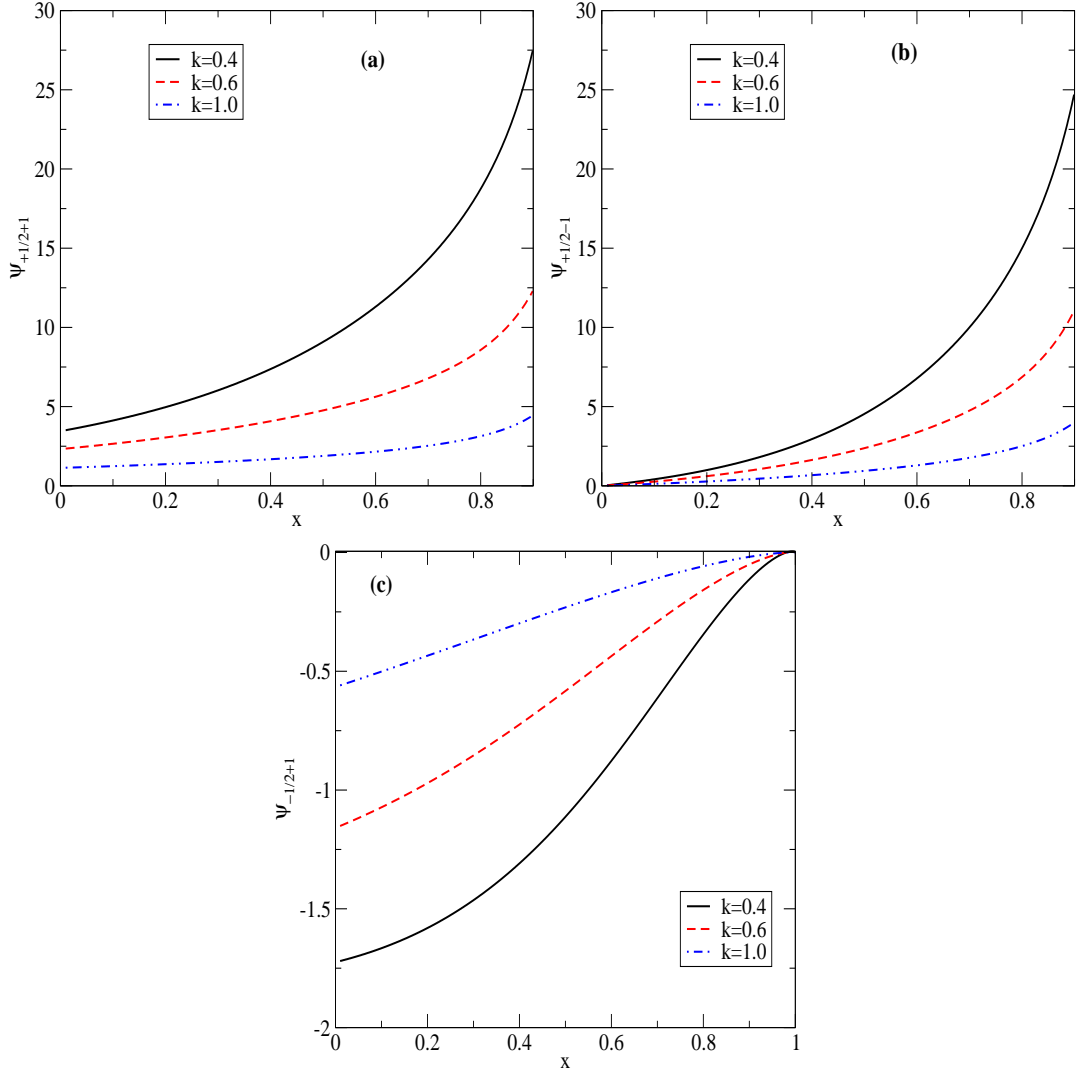

\centering
\begin{subfigure}{
\begin{minipage}[c]{0.9\textwidth}
\centering
\includegraphics[width=7cm,height=7cm,clip]{longnew_fig2a.eps}%
\hspace{0.2cm}%
\includegraphics[width=7cm,height=7cm,clip]{longnew_fig2b.eps}
\end{minipage}%
}
\end{subfigure}
\centering
\begin{minipage}[c]{0.5\textwidth}
\centering
\includegraphics[width=7cm,height=7cm,clip]{longnew_fig2c.eps}
\end{minipage}%
\caption{\label{fig1} (Color online) Two-particle LFWFs of the electron vs. $x$ for
$M=0.51$ MeV, $m=0.5$ MeV, $\lambda=0.02 $ MeV and fixed values of
$\mid k^\perp \mid =k $ in units of MeV. In (b) and (c) we have
divided the LFWFS by the factors $(k^1+i k^2)$ and $(-k^1+i k^2)$
respectively. }
\end{figure}

All Fourier transforms have been performed by numerically calculating the
Fourier sine and cosine transforms and then calculating the resultant by
squaring them, adding and taking the square root, thereby yielding the
Fourier Spectrum.  The amplitude is divided
by the normalization factor ${e^4\over 16 \pi^3}$. In Fig. \ref{fig1},
we have shown
the two particle LFWFs of the electron as a function of $x$ for different
$k^\perp$.
We have taken $m=0.5$ MeV, $M=0.51$ MeV and $\lambda = 0.02$ MeV. The wave
functions are similar for a slight change of parameter values, however for
$m<M$, there will be a node in $\psi_{-1/2+1}(x, k^\perp)$ at $m=x M$, which
is seen in Fig. 1 (c). The effect of the node is almost negligible for these
parameter values.

\begin{figure}
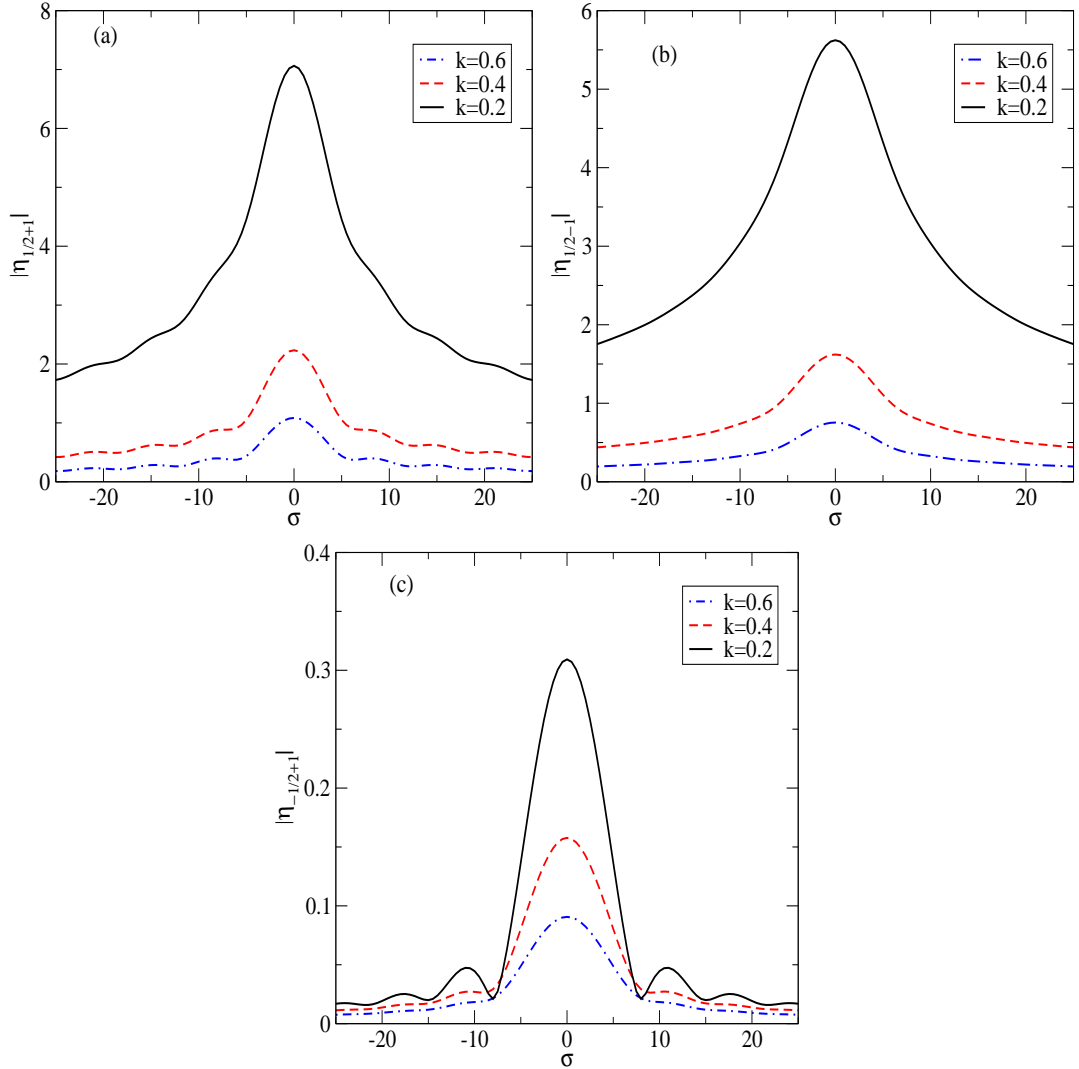

\centering
\begin{subfigure}{
\begin{minipage}[c]{0.9\textwidth}
\centering
\includegraphics[width=7cm,height=7cm,clip]{longnew_fig3a.eps}%
\hspace{0.2cm}%
\includegraphics[width=7cm,height=7cm,clip]{longnew_fig3b.eps}
\end{minipage}%
}
\end{subfigure}
\centering
\begin{minipage}[c]{0.5\textwidth}
\centering
\includegraphics[width=7cm,height=7cm,clip]{longnew_fig3c.eps}
\end{minipage}%

\caption{\label{fig2} (Color online) Fourier spectrum of the two-particle LFWFs of
the electron vs. $\sigma$  for $M=0.51$ MeV, $m=0.5$ MeV,
$\lambda=0.02$ MeV and fixed values of $\mid k^\perp \mid =k $ in MeV. In
(b) and (c) we have divided the LFWFS by the factors $(k^1+i k^2)$
and $(-k^1+i k^2)$ respectively. }
\end{figure}

In Fig. \ref{fig2} we have shown the Fourier  Spectrum (FS) of
the 2-particle
LFWFs given by Eqs. (\ref{vsn2}), for the same mass parameters as in
Fig. \ref{fig1}. The Fourier transform (FT) has been
taken with respect to $x$ for fixed values of transverse momentum $k^\perp$.
The wave functions  $\eta_{1/2+1}$,
$\eta_{-1/2+1}$ and $\eta_{1/2-1}$ are obtained as,
\be
\eta^\uparrow_{1/2+1}(\sigma, k^\perp) &=& {1\over {2 \pi(-k^1+i k^2)}}
\int_0^1
dx~ e^{{i} \sigma
(x-{\hat x}_k)}~ \psi^{\uparrow}_{1/2+1} (x, k^\perp),\nonumber\\
\eta^\uparrow_{1/2-1}(\sigma, k^\perp) &=& {1\over
{2 \pi (k^1+i k^2)}} \int_0^1 dx
~e^{{i} \sigma
(x-{\hat x}_k)}~ \psi^{\uparrow}_{1/2-1} (x, k^\perp),\nonumber\\
\eta^\uparrow_{-1/2+1}(\sigma, k^\perp) &=& {1\over 2 \pi} \int_0^1 dx
~e^{{i} \sigma (x-{\hat x}_k)}~ \psi^{\uparrow}_{-1/2+1} (x, k^\perp),
\ee
where ${\hat x}_k={\sqrt{m^2+k_\perp^2} \over \sum_i\sqrt{m_i^2+k_{\perp i}^2}}$
is the peak of the distribution - where all the constituents in the n-particle
 Fock state have equal rapidity.

\begin{figure}
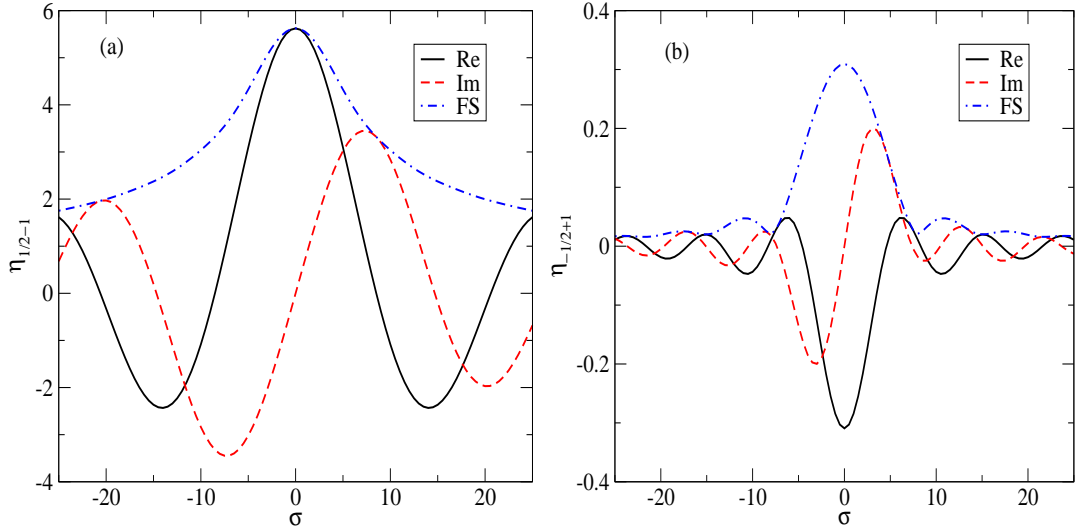

%\centering
\begin{minipage}[t]{0.9\textwidth}
\centering
\includegraphics[width=7cm,height=7cm,clip]{longnew_fig4a.eps}%
\hspace{0.2cm}%
\includegraphics[width=7cm,height=7cm,clip]{longnew_fig4b.eps}
\end{minipage}%
\caption{\label{fig2d} (Color online) Fourier transform (FT) of the two particle
LFWFs of the electron vs. $\sigma$  for $M=0.51$ MeV, $m=0.5$ MeV,
$\lambda=0.02$ MeV and fixed values of $\mid k^\perp \mid =k $ in
MeV. Re and Im denote the real and imaginary parts of the FT, and FS
denotes the Fourier spectrum presented in Figs. \ref{fig2}(b) and
(c). }
\end{figure}

%In the plots, the wave function $\eta$ is  multiplied by the normalization
%factor $2\pi$.
All helicity components of the wave function show peaks at
$\sigma=0$, the height of the peak sharply increases as $k^\perp$
decreases and decays away from $\sigma=0$.
In Fig. \ref{fig2d}, we have shown the complete Fourier transforms of the two
 particle LFWF for $k$=0.2 corresponding to the FS presented in
 Fig. \ref{fig2}(b)
and (c) to illustrate the difference. Though the real and imaginary parts of
the FT (i.e., the cosine and sine transforms, respectively) individually
exhibit a diffraction pattern, in Fig. \ref{fig2d}(a) they are just out of phase
to produce
any diffraction pattern in the FS. It is well-known in the theory of the
Fourier representation of signals \cite{oppenheim-lim} that the amplitude
and phase play different roles and in some cases many of the important
features of a signal are preserved only if the phase is retained.

\begin{figure}
%\centering
\begin{minipage}[t]{0.9\textwidth}
\centering
\includegraphics[width=7cm,height=7cm,clip]{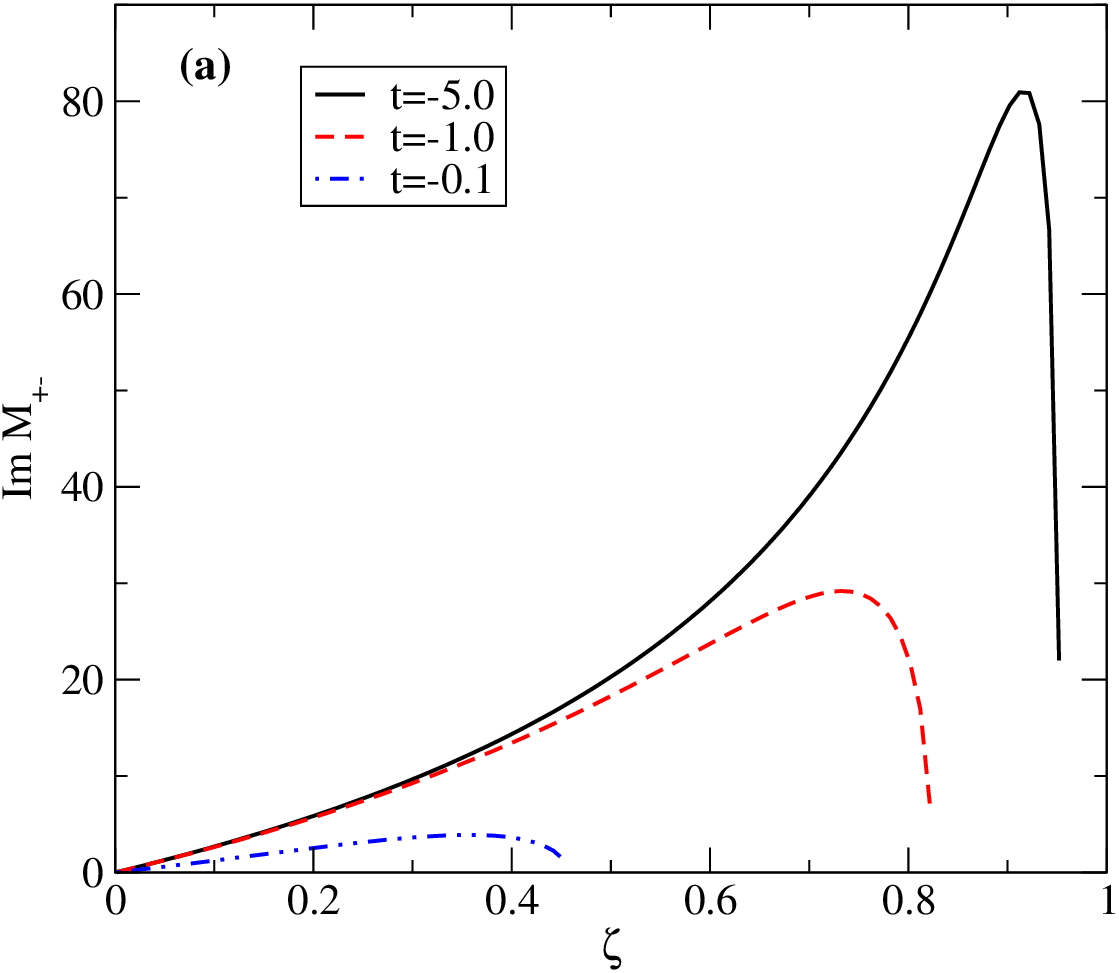}%
\hspace{0.2cm}%
\includegraphics[width=7cm,height=7cm,clip]{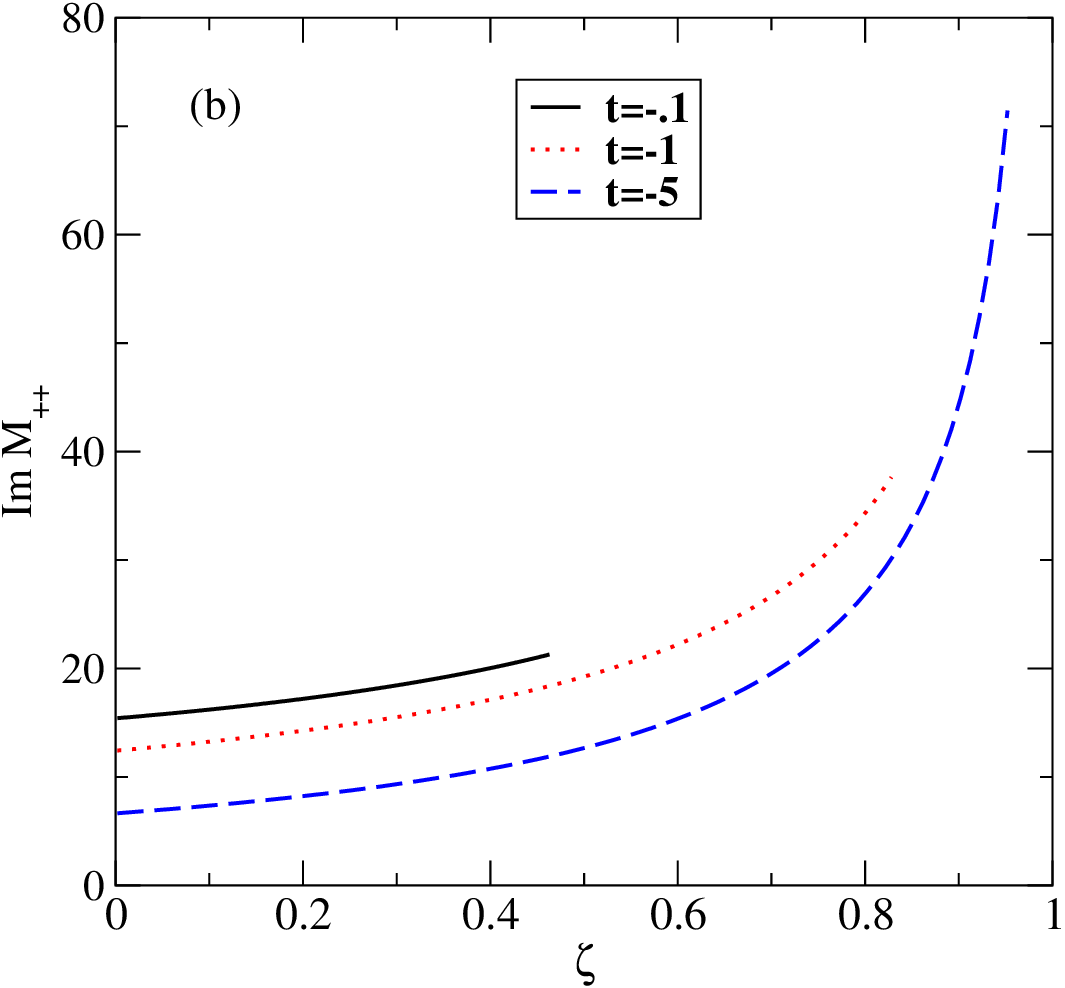}
\end{minipage}%
\caption{\label{fig3} (Color online) Imaginary part of the DVCS amplitude for an
electron vs. $\zeta$  for different values of $t$ : (a) helicity-flip
part, (b) helicity non-flip part. We have taken $M=0.51$ MeV,
$m=0.5$ MeV, $\lambda=0.02$ MeV. The parameter $t$ is given in MeV.}
\end{figure}

\begin{figure}
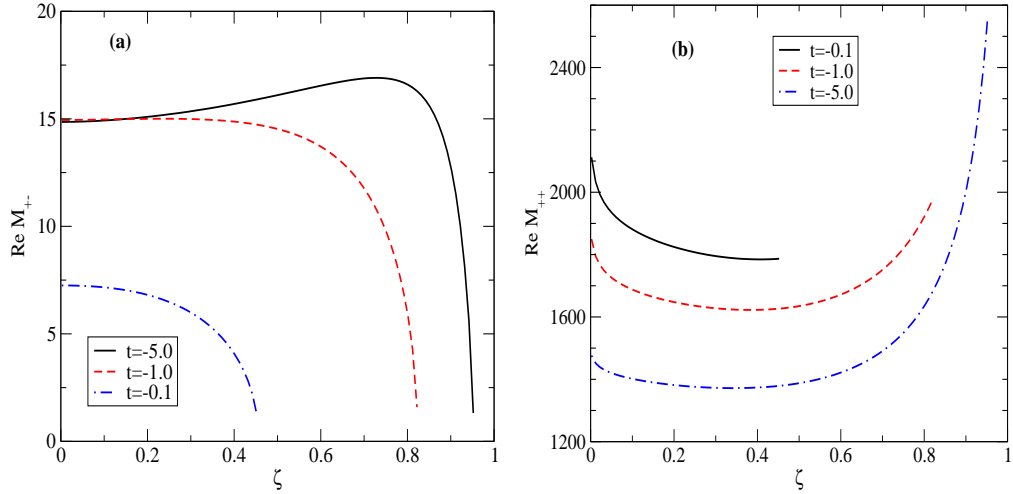

\centering
%\begin{minipage}[c]{0.9\textwidth}
\centering
\includegraphics[width=6.5cm,height=6.5cm,clip]{longnew_fig6a.eps}
\hspace{0.2cm}%
\includegraphics[width=6.5cm,height=6.5cm,clip]{longnew_fig6b.eps}
%\end{minipage}%
\caption{\label{fig4} (Color online) Real part of the DVCS amplitude for an
electron vs. $\zeta$  for different values of $t$ : (a) helicity-flip
part, (b) helicity non-flip part. We have taken $M=0.51$ MeV,
$m=0.5$ MeV, $\lambda=0.02$ MeV. The parameter $t$ is given in MeV.}
\end{figure}

The plots of the DVCS amplitude have been done by fixing $-t$ and varying
both $\zeta$ and $\Delta_\perp$.
In Fig. \ref{fig3}(a) we have shown the imaginary part of the helicity flip
DVCS amplitude $M_{+-}$ as a function of $\zeta$ for different values
of $-t$.
% We have taken $m=0.5$ MeV, $M=0.51$ MeV and $\lambda=0.02$ MeV.
Im$[M_{+-}]$ is zero as $\zeta \rightarrow 0$,  increases
continually with $\zeta$, and then falls down sharply at the end. It
increases for higher $t$ for the same $\zeta$. Fig. \ref{fig3}(b)
shows the helicity non-flip part of the corresponding amplitude
Im$[M_{++}]$ vs. $\zeta$. Unlike Im$[M_{+-}]$, Im$[M_{++}]$ is
non-vanishing at $\zeta=0$, it  decreases for higher $-t$. The
largest allowed value of $\zeta$ is given by Eq. (\ref{nn4}) for
fixed $t$. Fig. \ref{fig4}(a) shows the plot of the real part of the
helicity flip DVCS amplitude for the same values of the parameters.
Re$[M_{+-}]$ is non-vanishing at $\zeta \rightarrow 0$. For small
$\zeta$, it is almost flat
  for a fixed $-t$ and then falls down at large  $\zeta$.
 Fig. \ref{fig4}(b) shows the plot of the real part of the corresponding
helicity
non-flip amplitude $M_{++}$ vs. $\zeta$. It shows a different functional
behavior as it increases both for small as well as large $\zeta$.

\begin{figure}
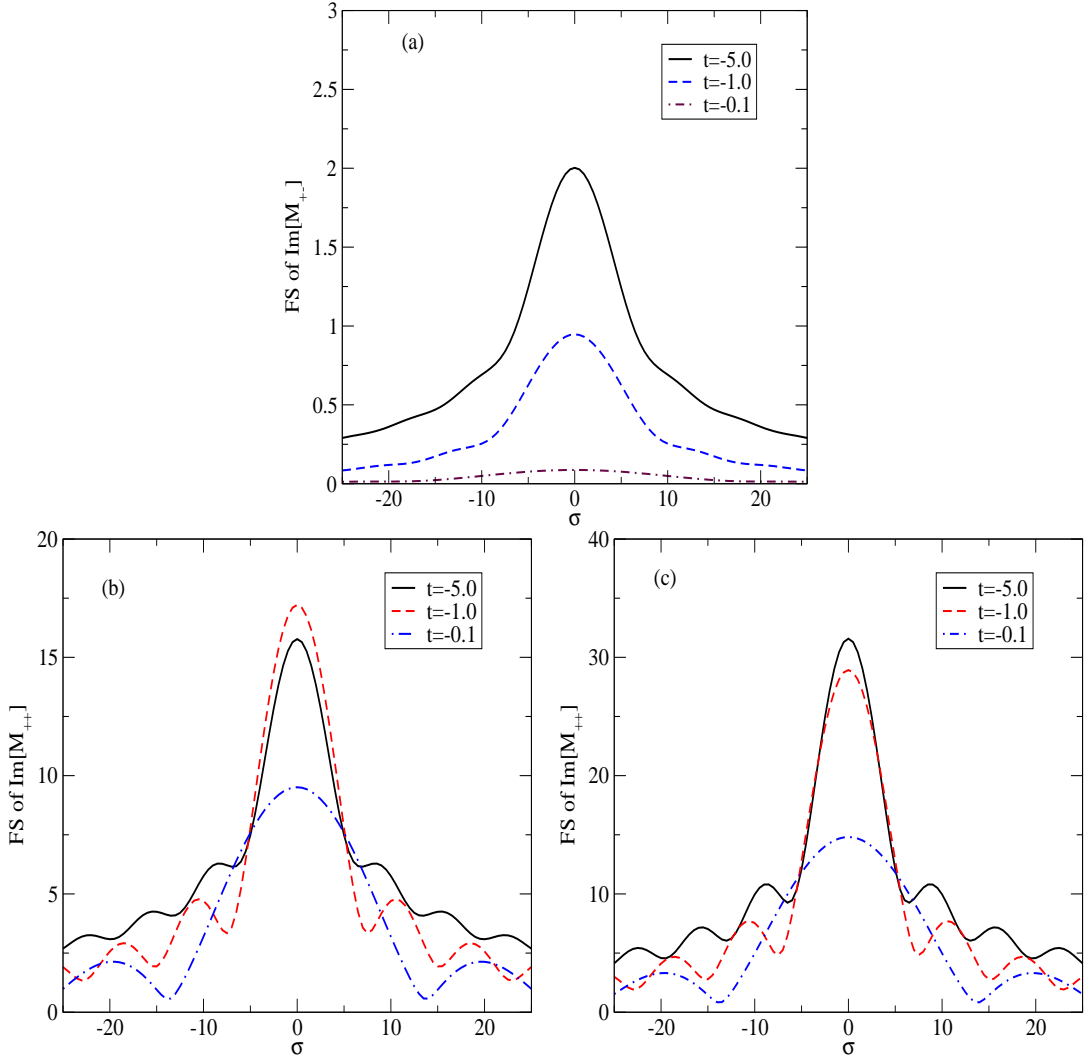

\centering
%\begin{minipage}[t]{0.5\textwidth}
\centering
\includegraphics[width=7cm,height=7cm,clip]{longnew_fig7a.eps}
%\hspace{0.2cm}%
%\includegraphics[width=7cm,height=7cm,clip]{longnew_fig7b.eps}
%\end{minipage}%
%\end{figure}
%\begin{figure}
\centering
\begin{minipage}[c]{0.9\textwidth}
%\centering
\includegraphics[width=7cm,height=7cm,clip]{longnew_fig7b.eps}
\hspace{0.2cm}%
\includegraphics[width=7cm,height=7cm,clip]{longnew_fig7c.eps}
\end{minipage}%
\caption{\label{fig5} (Color online) Fourier spectrum of the imaginary part of the
DVCS amplitude for an electron vs. $\sigma$  for different values
of $t$ : (a) when the electron helicity is flipped, (b) and (c) when
the helicity is not flipped. In (b) $Q=10$ MeV, in (c) $Q=50$ MeV.
The mass parameters are $M=0.51$ MeV,
$m=0.5$ MeV, $\lambda=0.02$ MeV.
 The parameter $t$ is given in MeV.}
\end{figure}

\begin{figure}
\centering
\includegraphics[width=7cm,height=7cm,clip]{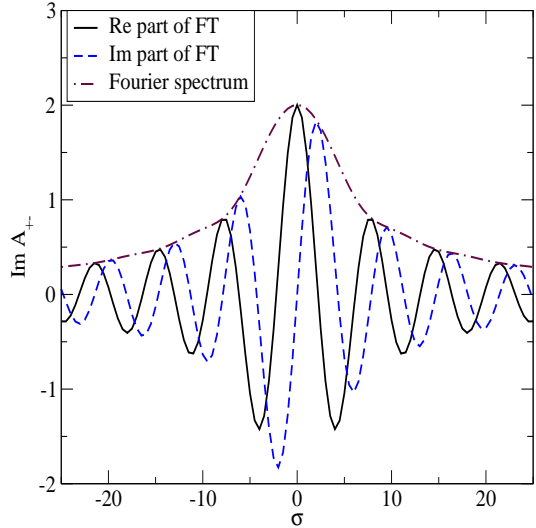}
\caption{\label{fig5d} (Color online) Fourier transform (FT) of the imaginary part of the
helicity flip DVCS amplitude for an electron vs. $\sigma$  for $t$=-5.0.
Re and Im denote the real and imaginary parts of the FT.
The mass parameters are $M=0.51$ MeV,
$m=0.5$ MeV, $\lambda$=0.02 MeV.}
\end{figure}

Fig. \ref{fig5} (a) shows the FS of the imaginary part of the
helicity flip amplitude vs. $\sigma$ for $M=0.51$ MeV, $m=0.5$ MeV
and $\lambda = 0.02$ MeV. The peak of the FS of Im[$M_{+-}$]
(i.e.,$|Im [A_{+-}]|$) increases with $\mid t \mid$.
%This is because the helicity
%flip DVCS amplitude is related to the Pauli form factor of the
%electron which falls with $\mid t \mid $ for large $\mid t \mid$.
The increasing behavior of the helicity flip amplitude $\gamma^*
e^\uparrow(S^e_z= 1/2) \to \gamma e^\downarrow(S^e_z= -1/2)$ at
small $\mid t \mid $ reflects the fact that one needs to transfer
one unit of orbital angular momentum $\Delta L_z = \pm 1$  to the
electron to conserve $J_z$. Note that the initial and final photon
are taken to have transverse polarization. A similar behavior is
expected for $\gamma^* p^\uparrow(S^p_z= 1/2) \to \gamma
p^\downarrow(S^p_z= -1/2)$ amplitude. The FS of Im$[M_{+-}]$ does
not show a diffraction pattern in $\sigma$. In Fig. \ref{fig5} (b)
and (c) we have shown the FS of the imaginary part of the helicity
non-flip DVCS amplitude vs. $\sigma$ for the same parameter values.
The helicity non-flip amplitude depends on the scale $\Lambda$. We
have taken $\Lambda=Q$. Fig. \ref{fig5} (b) shows the plot for
$Q=10$ MeV, 5(c) is for $Q=50$ MeV. Im $[A_{++}]$ shows a
diffraction pattern, the peak becomes narrower  as $\mid t \mid $
increases.
% The different
%behavior of the Im$[A_{+-}]$ and $|Im[A_{++}]|$, namely, the absence of
%diffraction pattern in Im$[A_{+-}]$ is due to the different behavior of
%Im$[M_{+-}]$ and  Im$[M_{++}]$ in $\zeta$, as seen in Fig. 3.
In Fig. \ref{fig5d} we have shown the complete Fourier transform of the
 imaginary part of the DVCS amplitude corresponding to the FS presented in
 Fig. \ref{fig5}(a). It again shows that the real (cosine transform) and the
 imaginary (sine transform) of the Fourier transform
individually show a diffraction pattern, but they are out of phase
and thus the FS does not show the diffraction pattern.

The number of minima in the diffraction pattern increases
with $\mid t \mid $ for fixed $Q$ or, in other words, the first minima
move in with increase of $\mid t \mid $. For $Q=50$ MeV,
the behavior is the same, the number of minima are higher for higher $t$
for the same $\sigma$ range. The number and the positions of the minima
are independent of $Q$. Only the magnitude of the peak changes with $Q$.
%The amplitude increases as $\mid
%t \mid$ increases for fixed $Q$, which is expected,  as the first
%moment with respect to $x$ of the helicity non-flip GPDs gives the
%Dirac form factor of the electron which increases logarithmically
%with $\mid t \mid $. The second moment is related to  the lepton
%contribution to the graviton-electron-electron form factor $A(t)$
%which has a similar behavior in $O(\alpha)$. We have checked that
%for $t << m^2$, the qualitative behavior of Im$[A_{++}]$ is the
%same and independent of $Q$, only the magnitude increases as $Q$
%increases.
 Some of the plots of the FS
of the DVCS amplitude show similarities with the FS of the LFWFs
themselves. The generalized parton distributions are related to
the form factors, and the form factors can be written
as overlaps of LFWFs. In fact, for a meson in $1+1$ dimensional QCD, the
form factor becomes an overlap of the LFWFs with different longitudinal
momentum fractions, $x$ \cite{ein}, and the contribution is similar to the
$2 \rightarrow 2$ part of DVCS amplitude.
\begin{figure}
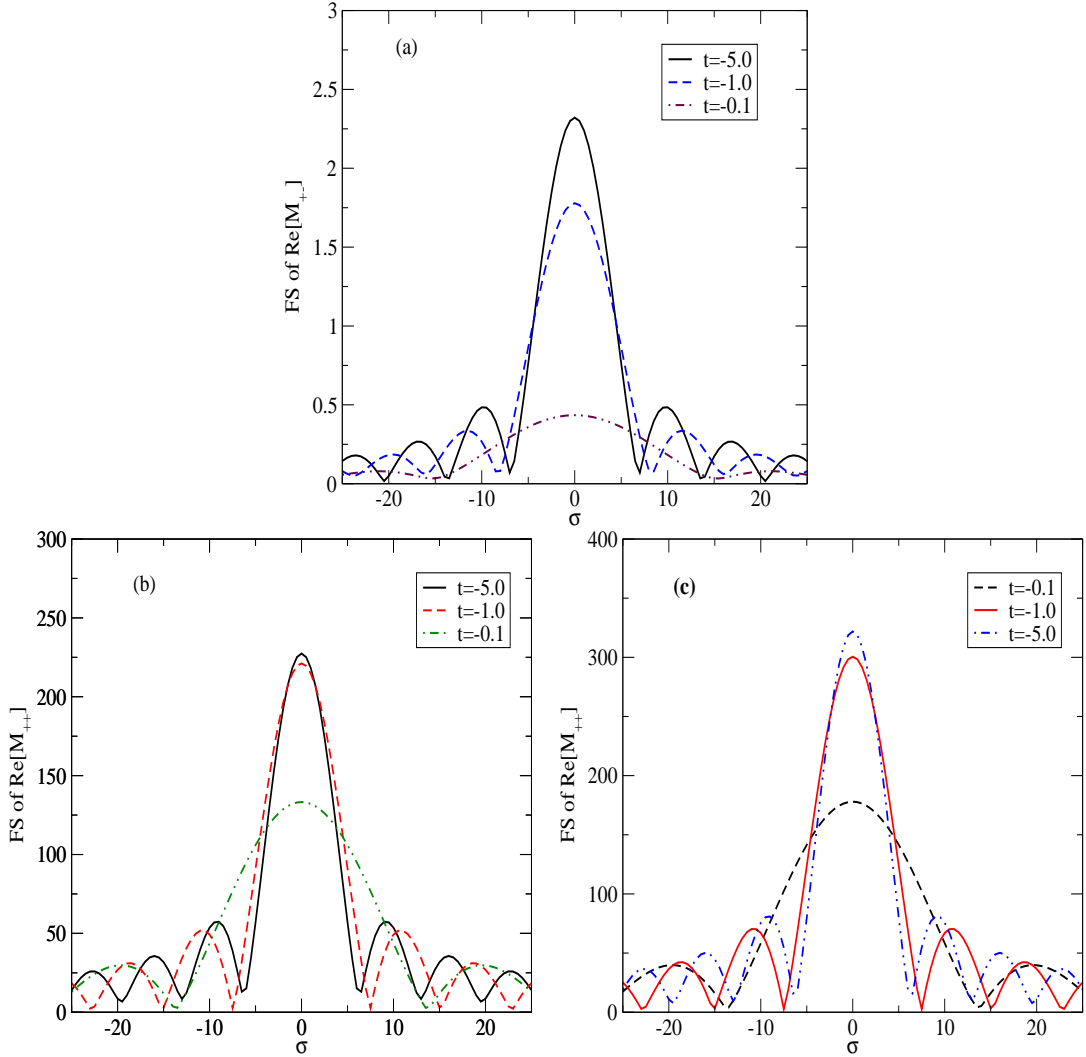

\centering
%\begin{minipage}[t]{0.5\textwidth}
%\centering
%\psfrag{a}{b}
%\psfrag{b}{a}
\includegraphics[width=7cm,height=7cm,clip]{longnew_fig9a.eps}
%\end{minipage}%
%\end{figure}
%\begin{figure}
\centering
\begin{minipage}[c]{0.9\textwidth}
%\centering
\includegraphics[width=7cm,height=7cm,clip]{longnew_fig9b.eps}
\hspace{0.2cm}%
\includegraphics[width=7cm,height=7cm,clip]{longnew_fig9c.eps}
\end{minipage}%
\caption{\label{fig6} (Color online) Fourier spectrum of the real part of the
DVCS amplitude for an electron vs. $\sigma$  for different values
of $t$ : (a) when the electron helicity is  flipped,
(b) and (c) when the helicity is not flipped. In (b) $Q=10$ MeV,
in (c) $Q=50$ MeV. The mass parameters are $M=0.51$ MeV,
$m=0.5$ MeV, $\lambda=0.02$ MeV. The parameter $t$ is given in MeV.}
\end{figure}

Fig. \ref{fig6} (a) shows the FS of the real part  of the
helicity flip amplitude vs. $\sigma$, where we chose the same
values of the parameters,  $M=0.51$ MeV, $m=0.5$ MeV and $\lambda
= 0.02$ MeV. The FS i.e.,
$|Re [A_{+-}]|$ shows a diffraction pattern in $\sigma$.
With the increase of $-t$, the central peak increases and
its width decreases .
The helicity flip part of the DVCS amplitude does not
depend on $Q$.

 The non-existence of the diffraction pattern in
 the FS of the imaginary part of the helicity flip amplitude
in $\sigma$ is due to its different behavior in $\zeta$, as seen
from Figs. \ref{fig3} and \ref{fig4}.
$Im [M_{+-}]$ decreases smoothly as $\zeta$ decreases  to vanish
 at $\zeta=0$ which is distinctly different from all other amplitudes including
$Re[M_{+-}]$. All other amplitudes show some flatness or a plateau in $\zeta$
and their FS in $\sigma$ space shows a diffraction pattern.

%%%%%%%%%%%%%%%%%%%%%%%%%%%%%%%%%%%%%%%%%%%%%%%%%%%%%%%%%%%%%%%%%%%%%%%%%%%
%We have found that the Fourier transforms of the real part of DVCS
%amplitudes with only  $3 \to 1$ contribution also show
%diffraction patterns, the corresponding functions
%show a plateau in $\zeta$ space before
%rapidly going to zero at $\zeta=0$. This step and plateau-like behavior is
%absent in $Im [M_{+-}]$ and it shows no diffraction pattern in
%the $\sigma$ space.
%%%%%%%%%%%%%%%%%%%%%%%%%%%%%%%%%%%%%%%%%%%%%%%%%%%%%%%%%%%%%%%%%%%%%%%%%%%%%

\begin{figure}
%\centering
\begin{minipage}[t]{0.5\textwidth}
\centering
\includegraphics[width=7cm,height=7cm,clip]{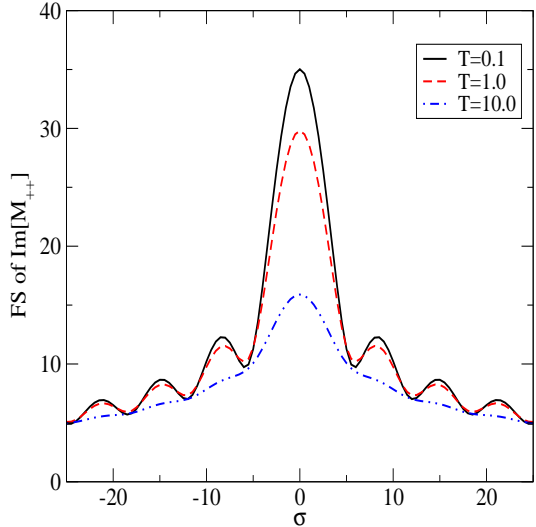}%
\end{minipage}%
\caption{\label{fig7} (Color online) Imaginary part of the helicity non-flip DVCS
amplitude for an electron vs. $\sigma$ for fixed $T$ in $MeV$.}
\end{figure}

\begin{figure}
%\centering
\begin{minipage}[t]{0.9\textwidth}
\centering
\includegraphics[width=7cm,height=7cm,clip]{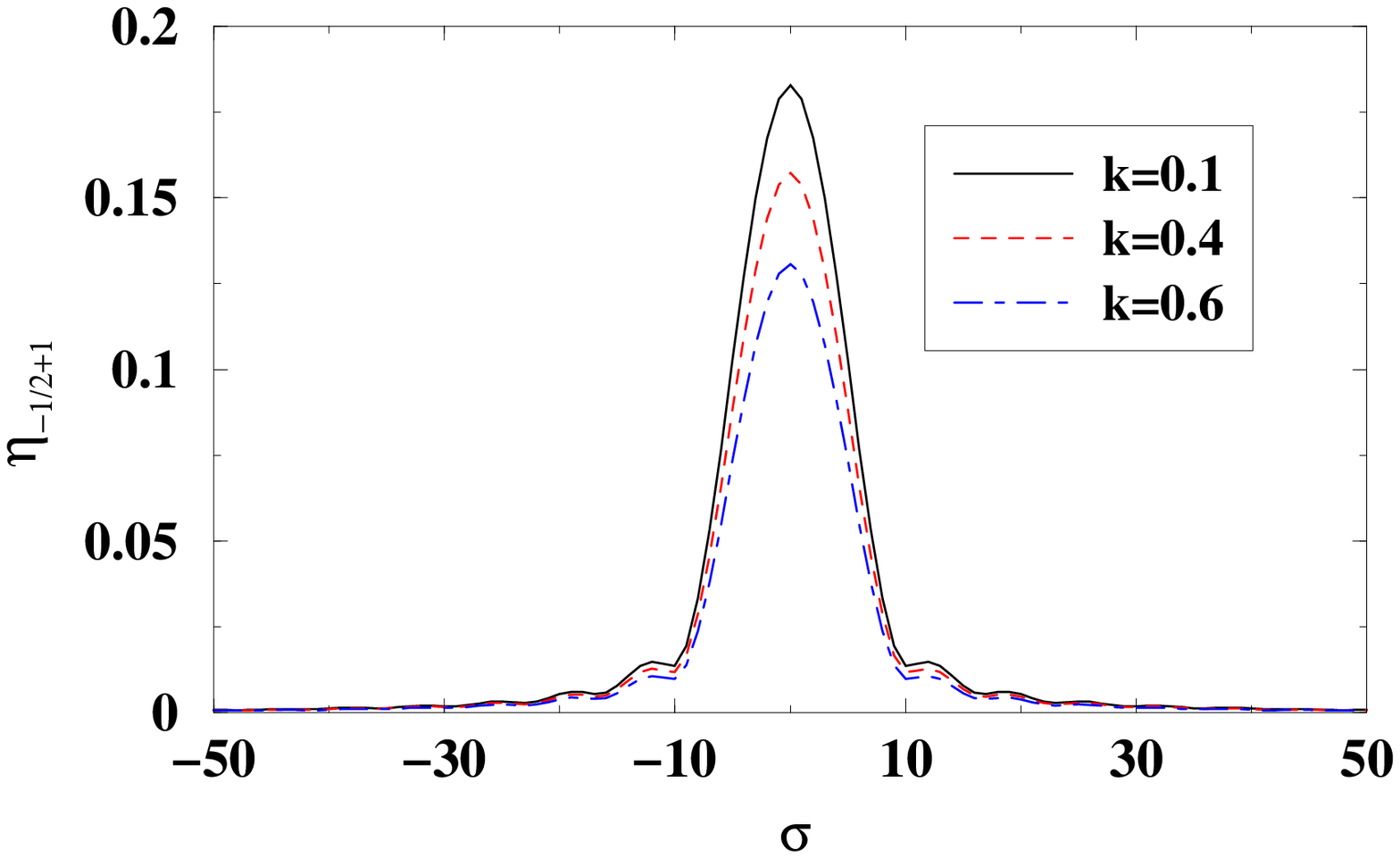}%
\hspace{0.2cm}%
\includegraphics[width=7cm,height=6.7cm,clip]{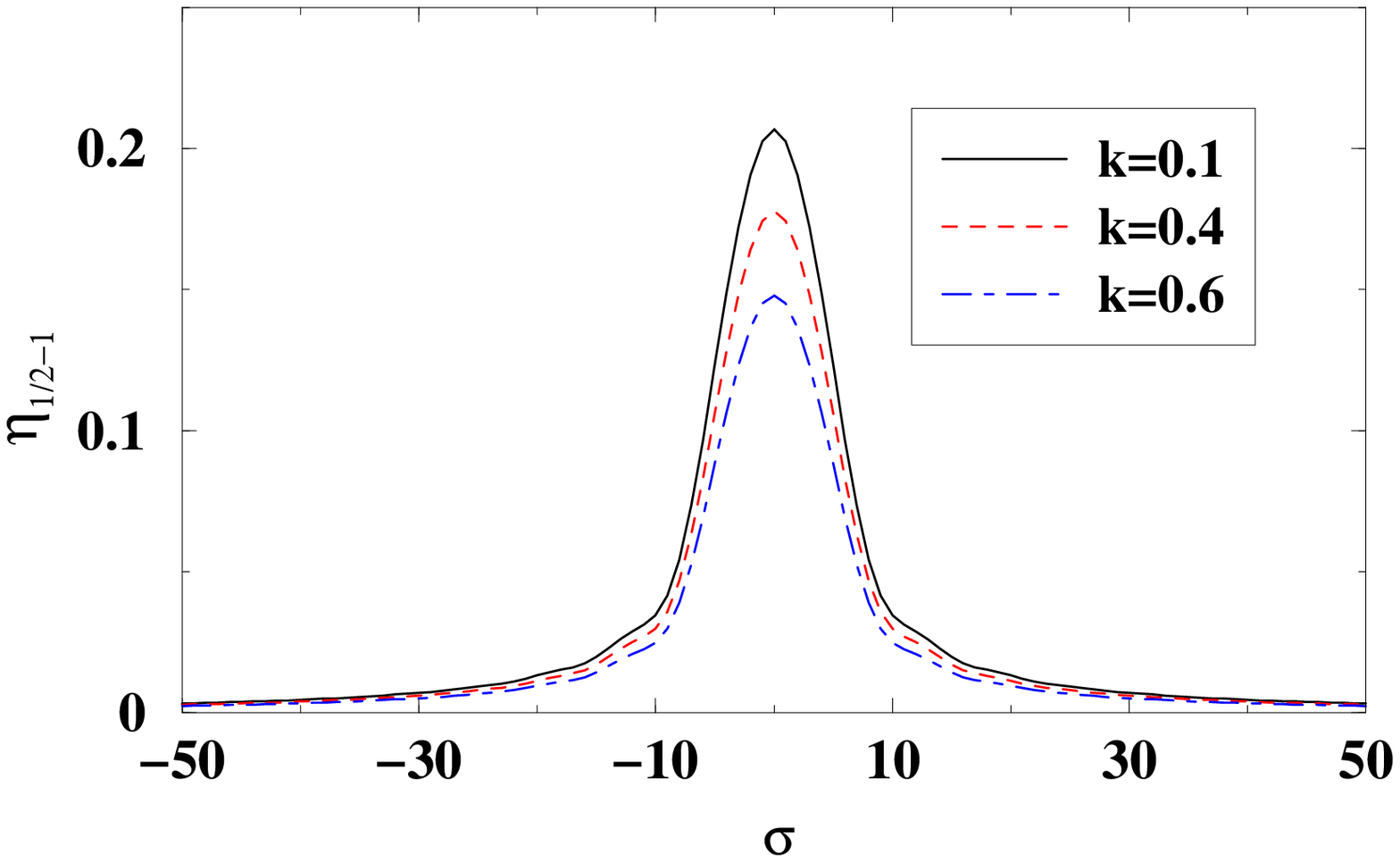}
%\end{minipage}%
%\end{figure}
%\begin{figure}
\centering
%\begin{minipage}[c]{0.5\textwidth}
%\centering
\includegraphics[width=7cm,height=7cm,clip]{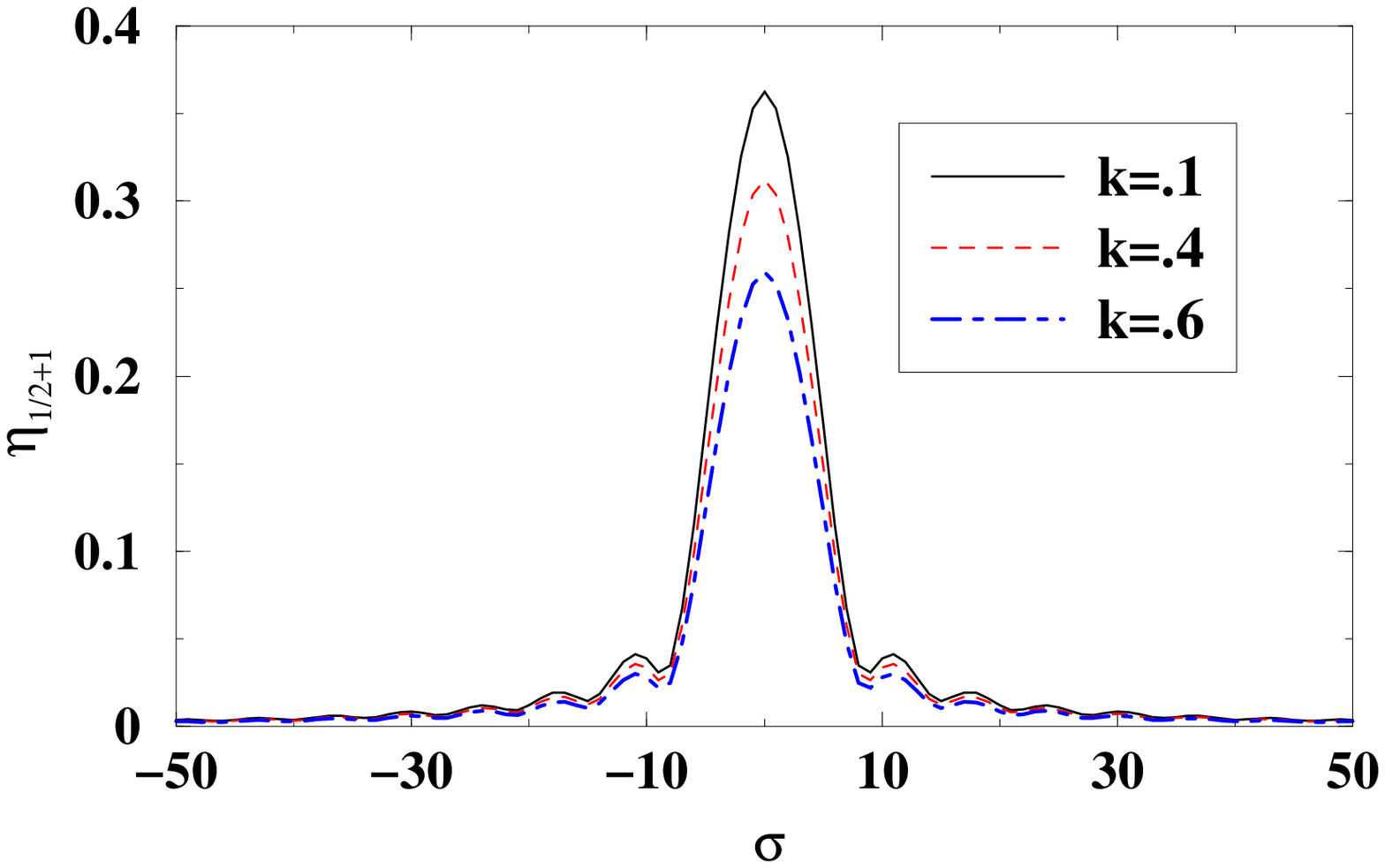}
\end{minipage}%
\caption{\label{fig8} (Color online) Fourier spectrum of the LFWFs of the simulated
hadron model vs. $\sigma$  for $M=0.5$ MeV, $m=\lambda=1.0$ MeV and
fixed values of $\mid k^\perp \mid =k $ in MeV. We have divided the
LFWFs by the normalization constant. In (b) and (c) we have divided
by the factors $(k^1+i k^2)$ and $(-k^1+i k^2)$ respectively as
well. }
\end{figure}

In Fig. \ref{fig6} (b) and (c), we have plotted the FS of the real part of
the
helicity non-flip DVCS amplitude vs. $\sigma$.
 One can see the diffraction pattern
here as well.
Fig. \ref{fig6}(b) is for $Q=10$ MeV and (c) is for $Q=50$ MeV. As before, the
 qualitative behaviors of the diffraction pattern do not change with $Q$. For
the same $\mid t \mid$, the number of minima and their positions are
independent of $Q$ for any fixed  $\mid t \mid$, only the height changes with
 $Q$. For each $Q$,
the peak at $\sigma=0$ is sharper and higher as $\mid t \mid $ increases.

Instead of using the $\zeta$ and $t$ variables, we can define another set of
variables $\zeta$ and $T$, where $T$ is defined as $T={\Big
({\Delta^\perp\over 1- \zeta}\Big )}^2$. The arguments of the final state
LFWF then are $x'={x-\zeta\over 1-\zeta}$ and $k'^\perp=k^\perp-(1-x)
\sqrt{T}$,
in other words, the transverse momenta become decoupled from $\zeta$. We
can now take $\zeta$ and $T$ as independent variables and the GPDs as well
as the DVCS amplitude can be expressed in terms of them. They are however,
connected through
\be
t=-{\zeta^2 M^2\over 1-\zeta}-(1-\zeta) T.
\ee
This relation determines the range of allowed values of $\zeta$ and $T$, such
that $t << Q^2$. In practice, $\zeta$ can never become very close to $1$.
The $\zeta$ dependence of the DVCS amplitude now comes purely from $x'$
for fixed $T$.
Fig. \ref{fig7} shows a plot of the imaginary part of the helicity
non-flip amplitude
for fixed $T$ vs. $\sigma$. For very small $\sigma$, the slope of the $\sigma$
distribution is given by,
\be
{d\over d \sigma} {\mathrm Im} A_{++} \rightarrow \int_\eps^{1- \eps}
dx x ( F^{22}_{++}(x,x,T)+F^{31}_{++}(x,x,T) ).
\ee
Thus the slope and therefore the width of the $\sigma$ distribution depends
on
the second moment of the GPDs at $x=\zeta$.

%%%%%%%%%%%%%%%%%%%%%%%%%%%%%%%%%%%%%%%%%%%%%%%%%%%%%%%%%%%%%%%%%%%%%%%%%%%%
\section{Simulated Bound States}
%%%%%%%%%%%%%%%%%%%%%%%%%%%%%%%%%%%%%%%%%%%%%%%%%%%%%%%%%%%%%%%%%%%%%%%%%

For the dressed electron state, the real part of the DVCS amplitude
depends on the cutoffs at $x=0$ and $x=1$. We have chosen the cutoff
scheme discussed earlier. The cutoff at $x=0$ is taken for the
numerical calculation and its effect on the result is small.
However, starting from this QED point-like model where the electron
fluctuates to spin-half plus spin one constituents, one can
construct LFWFs for the hadrons. In the two- and three-particle
LFWFs for the electron, the bound state mass $M$ appears in the
energy denominators. A differentiation of the QED LFWFs with respect
to  $M^2$  improves the convergence at the end points: $x=0,1,$ as
well as at high $k^2_\perp$, thus simulating a bound state valence
wavefunction. Differentiating once with respect to $M^2$ will
generate a meson-like behavior of the LFWF. Thus we write the hadron
two-particle LFWFs as \be {\tilde
\psi}_{s_1s_2}(x,k_\perp)=M^2{\partial \over \partial M^2}
\psi_{s_1s_2}(x,k_\perp) \ee where $\psi_{s_1s_2}(x,k_\perp)$ are
the electron LFWFs. Taking the Fourier transform in $k_\perp$ also
we can write the wavefunctions in $\sigma$ and  transverse impact
parameter $b_\perp$ as \be \chi_{s_1s_2}(\sigma,b_\perp)={1\over
(2\pi)^3}\int_0^1 dx\int d^2k_\perp e^{i \sigma (x-{\hat x}_k)} e^{i
k_\perp \cdot b_\perp}  {\tilde \psi_{s_1s_2}}(x,k_\perp), \ee where
the peak of the distribution ${\hat x}_k= {\sqrt{m^2+k_\perp^2}\over
\sum_i{\sqrt{m_i^2+k_{\perp i}^2}}}$. Writing $k_\perp \cdot
b_\perp= k b \cos\theta$, where $b=|b_\perp|$ and $k=|k_\perp|$ and
performing the integration over $\theta$ we obtain \be
\chi_{s_1s_2}(\sigma,b_\perp)={1\over (2\pi)^3}\int_0^1 dx\int k dk
e^{i \sigma (x-{\hat x}_k)}  (2\pi) J_0(bk) {\tilde
\psi_{s_1s_2}}(x,k_\perp), \ee where $J_0(bk) $ is the Bessel
function. For the wavefunctions $\chi_{1/2 \pm 1}$ we have the
explicit momentum components $(k_1\pm i k_2)$ present in the
numerators. We use \be \int d^2k_\perp (k_1 \pm i k_2) e^{i k_\perp
\cdot b_\perp}&=&(-i)\int d^2k_\perp({\partial \over
\partial b_1}
\pm i {\partial \over \partial b_2})e^{i k_\perp \cdot b_\perp}\nonumber \\
%&=&(-i)\int d^2k_\perp{(b_1\pm i b_2)\over b}{\partial \over \partial
%b}e^{ik b \cos\theta} \nonumber\\
&=&(-i)\int k dk d\theta {k(b_1\pm i b_2)\over b}{\partial \over \partial (k
b)}e^{ik b \cos\theta}\nonumber\\
&=&(-i)\int k^2 dk {(b_1\pm i b_2)\over  b}{\partial \over \partial (k b)}(2
\pi)J_0(kb)\nonumber\\
&=& 2 \pi i \int k^2 dk {(b_1\pm i b_2)\over b}J_1(kb)
\ee

For  the plot of the wavefunctions we set $b_2=0$ in the above
expression, \be
 \chi_{+1/2+1}(\sigma,b) &=&{e\over (2 \pi)^2}\sqrt{2}M^2 i\int dx
\int k^2 dk {x
(1-x)^{1/2} J_1(kb)e^{i\sigma (x-{\hat x}_k)}
 \over (M^2 x(1-x)-k^2-m^2 (1-x)-\lambda^2 x)^2}\nonumber \\
 \chi_{+1/2-1}(\sigma,b)  &=&{e\over (2 \pi)^2}\sqrt{2}M^2 i\int dx
\int k^2 dk {x^2 (1-x)^{1/2} J_1(kb)e^{i\sigma (x-{\hat x}_k)}
 \over (M^2 x(1-x)-k^2-m^2 (1-x)-\lambda^2 x)^2}\nonumber \\
 \chi_{-1/2+1}(\sigma,b)  &=&{e\over (2 \pi)^2}\sqrt{2}M^2\int dx
\int k dk {x^2 (1-x)^{3/2} J_0(kb)e^{i\sigma (x-{\hat x}_k)}
 \over (M^2 x(1-x)-k^2-m^2 (1-x)-\lambda^2 x)^2}\nonumber \\
\label{psift}
\ee
For computational purpose, we use
\be
J_n(kb)={1\over \pi}\int_0^\pi d\theta \cos(n \theta-kb \sin\theta).
\ee
%The Fourier Spectra of the LFWFs of the hadron model given in Eq.
%(\ref{psift}) are shown in Fig. \ref{fig8ft}.
\begin{figure}
\begin{minipage}[t]{0.9\textwidth}
\centering
\psfrag{B}{$|b_\perp|$}
\psfrag{S}{$\sigma$}
\psfrag{P1}{$\chi_1$}
\psfrag{P22}{$\chi_2$}
\psfrag{P3}{$\chi_3$}
%\centering
\includegraphics[width=7cm,height=7cm,clip]{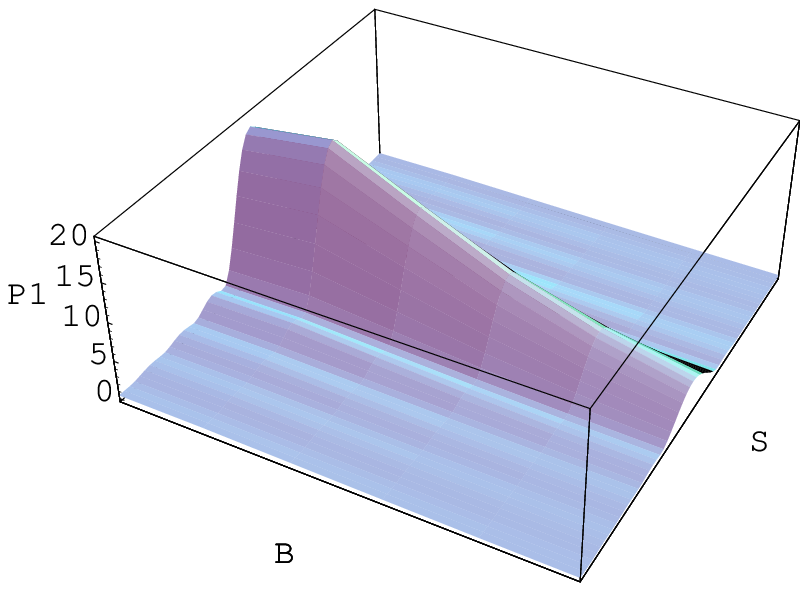}%
\hspace{0.2cm}%
\includegraphics[width=7cm,height=6.7cm,clip]{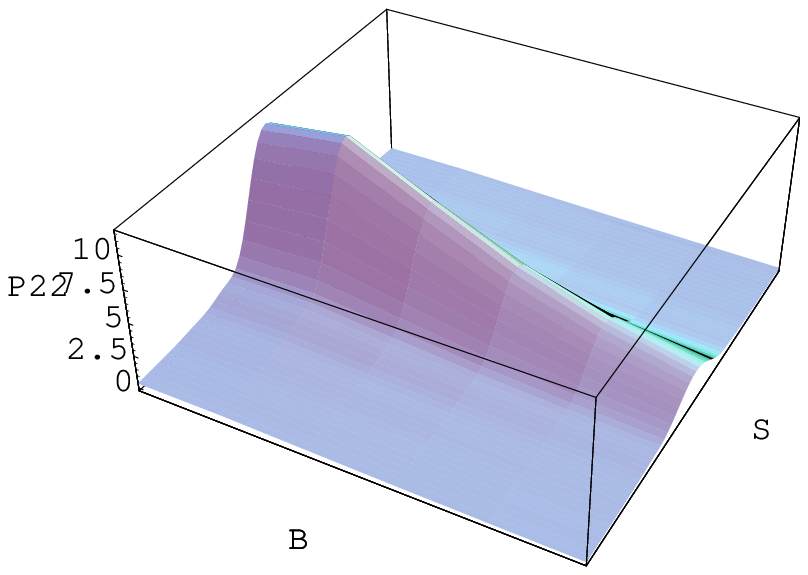}
%\end{minipage}%
%\end{figure}
%\begin{figure}
\centering
%\begin{minipage}[c]{0.5\textwidth}
%\centering
\includegraphics[width=7cm,height=7cm,clip]{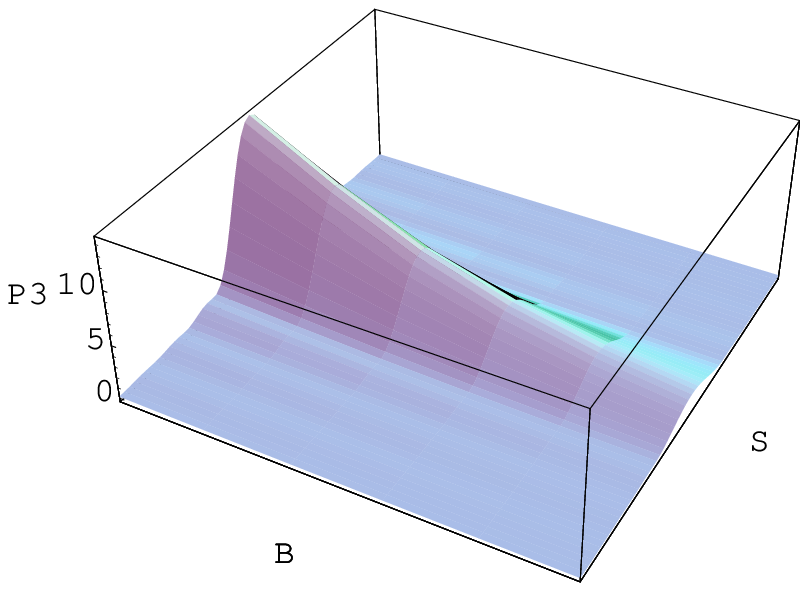}
\end{minipage}%
\caption{\label{fig8ft} (Color online) Fourier Spectrum of the LFWFs of the simulated
hadron model
plotted in $\sigma$, $|b_\perp|$ space  for $M=150$ MeV, $m=\lambda=300$ MeV.
$\chi_1$,$\chi_2$ and $\chi_3$ are $\mid\chi_{1/2+1}\mid$,
$\mid\chi_{1/2-1}\mid$ and
$\mid\chi_{-1/2+1}\mid$, respectively. In the plots $|b_\perp|$ runs
from 0.001 to 0.01 $\mathrm{MeV}^{-1}$ and $\sigma$ runs from -25 to +25. }
\end{figure}

\begin{figure}
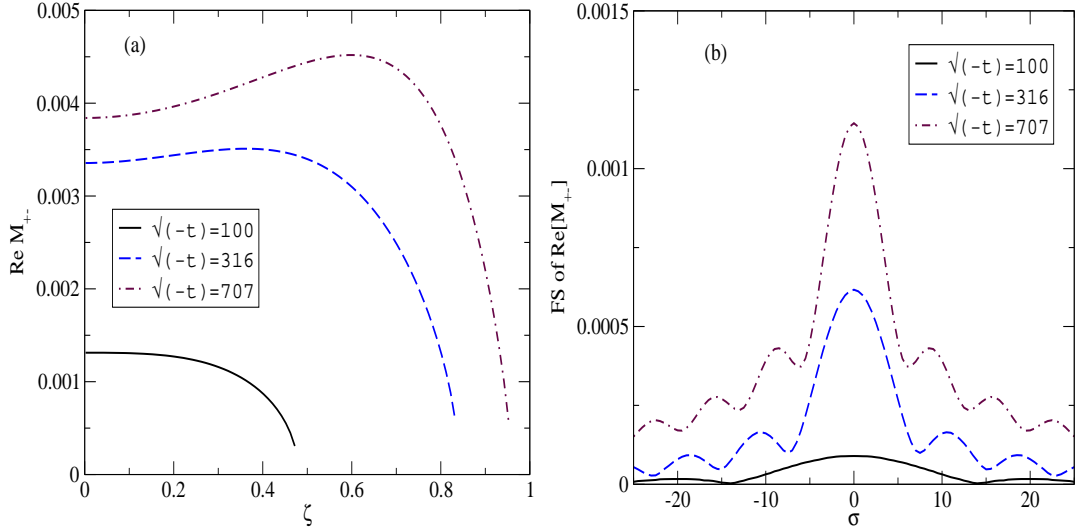

%\centering
\begin{minipage}[t]{0.9\textwidth}
\centering
\includegraphics[width=7cm,height=7cm,clip]{longnew_fig13a.eps}%
\hspace{0.2cm}%
\includegraphics[width=7cm,height=7cm,clip]{longnew_fig13b.eps}
\end{minipage}
\caption{\label{fig9} (Color online)  Real part of the DVCS amplitude for the
simulated hadron state.
The parameters are $M=150, m=\lambda=300$ MeV.
(a) Helicity flip amplitude  vs. $\zeta$, (b)
Fourier spectrum of the same vs. $\sigma$.
The parameter $t$ is in ${\mathrm{MeV}^2}$.}
\end{figure}

This procedure does not provide  an actual  model for a `meson'
wavefunction since the two constituents have spin-half and spin-one.
However, if we differentiate once more we can simulate the fall-off
at short distances which matches the fall-off wavefunction of a
baryon, in the sense that the form factor $F_1(Q^2)$ computed from
the Drell-Yan-West formula will fall-off like ${1\over Q^4}$. In
this case, we obtain a quark plus spin-one diquark model of a
baryon. Convolution of these wavefunctions in the same way as we
have done for the dressed electron wavefunctions will simulate the
corresponding DVCS amplitudes for bound-state hadrons. Note that the
differentiation of the single-particle LFWF will give a vanishing
result and as a result, the $3 \to 1$ contribution to the DVCS
amplitude vanishes in this model.  The resulting $\gamma^* p \to
\gamma p$ DVCS amplitude has both real \cite{real} and imaginary
parts \cite{imag}.

If we consider a dressed electron, the imaginary part from the pole
at $x = \zeta$ survives because of the numerator ${1\over x-\zeta}$
factor in the electron's LFWF. This numerator behavior reflects the
spin-1 nature of the constituent boson. The $x -\zeta \to 0$
singularity is shielded when we differentiate the final state  $n=2$
and $n= 3$ LFWFs with respect to $M^2$ and, as a result,  the
imaginary part of the amplitude vanishes in this model. We thus have
constructed a model where the DVCS amplitude is purely real.
However, the forward virtual Compton amplitude $\gamma^* p \to
\gamma^* p$ (whose imaginary part gives the structure function) does
not have this property.  The pole at $x = \zeta$ is not shielded
since the initial and final $n=2$ LFWFs are functions of $x$. It is
worthwhile to point out that in general the LFWFs for a hadron may
be non-vanishing at the end points \cite{dalley}, and recent
measurements of single spin asymmetries suggest that the GPDs are
non-vanishing at $x= \zeta$ \cite{exp}. A more realistic estimate
would require non-valence Fock states \cite{crji}.
\begin{figure}[t]
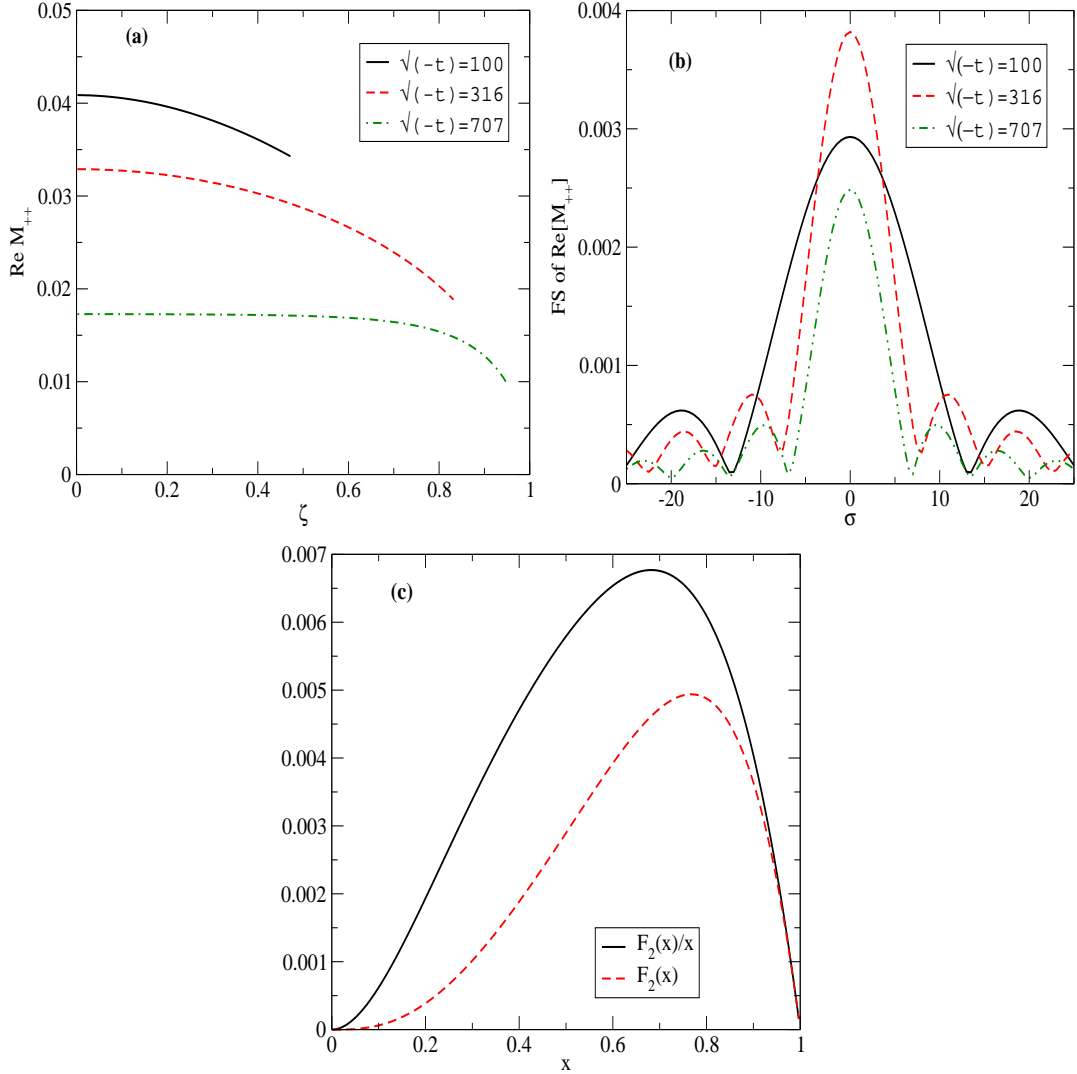

\centering
\includegraphics[width=7cm,height=7cm,clip]{longnew_fig14a.eps}%
\hspace{0.2cm}%
\includegraphics[width=7cm,height=7cm,clip]{longnew_fig14b.eps}
\begin{minipage}[c]{0.5\textwidth}
%\centering
\vspace{0.2cm}%
\includegraphics[width=7cm,height=7cm,clip]{longnew_fig14c.eps}
\end{minipage}%
\caption{\label{fig10} (Color online) Real part of the DVCS amplitude for the simulated
hadron bound state.
The parameters are $M=150, m=\lambda=300$ MeV.
(a) Helicity non-flip amplitude  vs. $\zeta$, (b)
Fourier spectrum of the same vs. $\sigma$, (c) Structure function vs. x.
The parameter $t$ is in ${\mathrm{MeV}^2}$.}
\end{figure}
An equivalent but easier way  to construct the hadronic model is to
differentiate the DVCS amplitudes with respect to the invariant mass
squared ($M^2$) of the initial and final bound states. Thus one can
calculate the quantity $M_F^2 {d\over dM_F^2} M_I^2 {d\over dM_I^2}
A_{ij}(M_I,M_F) $ where $M_I, M_F$ are the initial and final bound
state masses.
% and $i,j =+,-$.
%\begin{figure}
%\centering
%\begin{minipage}[t]{0.9\textwidth}
%\centering
%\includegraphics[width=7cm,height=7cm,clip]{dMhadFT_hf_m1M.5l1.eps}%
%\hspace{0.2cm}%
%\includegraphics[width=7cm,height=7cm,clip]{dMhadFT_hnf_m1M.5l1.eps}
%\caption{\label{fig10} (Color online) FT of the (a) Helicity flip and  (b) helicity
%non-flip DVCS amplitude in the simulated hadron model vs. $\sigma$.
%We have taken for $M=0.5$ MeV, $m=\lambda=1.0$ MeV. }
%\end{minipage}
%\end{figure}

For numerical computation we use the  discrete (in the sense that the
denominator is finite)  version of the
differentiation:
\be
M^2 {d A\over d M^2}= {\bar M^2} {A(M_1^2)-A(M_2^2) \over \delta M^2}
\ee
where ${\bar M^2}=(M_1^2+M_2^2)/2$ and $\delta M^2 = (M_1^2-M_2^2)$.
Then we have
\be
M_F^2 {d\over dM_F^2} M_I^2 {d\over dM_I^2} A_{ij}(M_I, M_F)
&=&{{\bar M_I^2}{\bar M_F^2}\over \delta M_I^2 \delta M_F^2} \big [
A_{ij}(M_{I1},M_{F1})-A_{ij}(M_{I1},M_{F2})\nonumber \\
&~&~~-A_{ij}(M_{I2},M_{F1})+
A_{ij}(M_{I2},M_{F2}) \big ];
\ee
$A_{ij}(M_I, M_F)$ is the DVCS amplitude for an electron.
The differentiation  with respect to $M_F^2$ of the amplitudes
(helicity flip and non-flip) brings in an extra
factor of $x-\zeta$ and thus the imaginary parts of the DVCS amplitudes
vanish in this model, as discussed before. The real parts of the
DVCS amplitudes survive and show diffraction patterns.
We take $M_{I1}, M_{F1} =150+1.0$ and  $M_{I2}, M_{F2} =150-1.0$
and the fixed parameters $M=150, m=\lambda=300$ MeV.
In Figs. \ref{fig8} and \ref{fig8ft} we have shown the FS of the
$2$-particle LFWFs for this model. The parameter values are scaled in
 Fig. \ref{fig8}, but the qualitative behaviours are the same.
Since  the wave function now vanishes at $x=0,1,$ the FS is
localized, and it decays sharply beyond $\mid \sigma \mid =10$. The
peak decreases more sharply for higher $k_\perp$ or lower
$|b_\perp|$. In Fig. \ref{fig9} (a) and (b) we have shown the
helicity-flip DVCS amplitude for the hadron model as functions of
$\zeta$ and its FS as a function of  $\sigma$ respectively. In Fig.
\ref{fig10} (a) and (b)  we have shown the real part of the helicity
non-flip amplitudes of the same model as a function of $\zeta$ and
its FS as a function of $\sigma$ respectively. Notice that the
helicity non-flip part of the amplitude no longer depends on the
scale. The amplitude decreases as $\zeta$ increases, in contrast to
the behavior for the electron. The FS  of both the helicity-flip and
non-flip DVCS amplitudes show a diffraction pattern in $\sigma$.
Fig. \ref{fig10}(c) illustrates the structure function $F_2(x)$ for
this model as  a function of $x$.

\subsection{An Optical Analog}

We propose an optics analog of the behavior of the FT  of DVCS
amplitude in $\sigma$. The similarity of  optics and  quantum fields
on the light cone  was first explored long ago  by Sudarshan, Simon
and Mukunda. They established the similarity of paraxial-wave optics
and light cone dynamics of  scalar \cite{sudarshan1} and  Maxwell
equations \cite{sudarshan2}.  In our case,  we are effectively
looking at the interference between the initial and final waves of
the scattered proton. The final-state proton wavefunction is
modified relative to the incident proton wavefunction because of the
momentum transferred to the quark in the hard Compton scattering.
The change in quark momentum along the longitudinal direction  can
be Fourier transformed to a shift in the light-front position of the
struck quark;  thus one can simulate a change in the quark's
longitudinal LF coordinate by an amount $\sigma= {1\over 2} b^-
P^+.$  This is analogous to diffractive scattering of a wave in
optics where $\sigma$ plays the role of the physical size of the
scattering center in a one-dimensional system. We are using $t$ to
register the change in the transverse momentum of the quark in the
scattering. The positions of the first minima move in with
increasing $\mid t \mid$.

Notice that the integrals over $x$ and $\zeta$ are of finite range.
More importantly, the upper limit of $\zeta$ integral is
$\zeta_{max}$ which in turn is determined by the value of $-t$. The
finiteness of slit width is a {\em necessary} condition for the
occurrence of diffraction pattern in optics. Thus when the
integration is performed over the  range $0$ to $\zeta_{max}$, the
finite range acts as a slit of finite width and provides a necessary
condition for the occurrence of diffraction pattern in the Fourier
transform of the DVCS amplitude.

When a diffraction pattern is produced, in analogy with single slit
diffraction, we expect the position of the first minimum to be inversely
proportional to $\zeta_{max}$. Since $\zeta_{max}$ increases with $-t$, we
expect the position of the first minimum to move to a smaller value of
$\sigma$, in analogy with optical diffraction. In the
case of the Fourier Spectrum of DVCS on the quantum
fluctuations of a lepton target in QED, and also in the
corresponding hadronic model, one sees that the diffractive
patterns in $\sigma$ sharpen and the positions of the first minima
typically move in with increasing momentum transfer. Thus the
invariant longitudinal size of the parton distribution becomes
longer and the shape of the conjugate light-cone momentum
distribution becomes narrower with increasing $\mid t \mid$.
%Regarding the diffraction patterns observed in the Fourier Spectrum
%of the DVCS amplitude, we further note that  for fixed $-t$, higher
% minima appear at
%positions which are integral multiples of the lowest minimum
% (as shown in Table \ref{tab1}. This
%further supports the  analogy with diffraction in optics.

\tablinesep=0.1in
\arraylinesep=0.1in
\extrarulesep=0.1in
\begin{table}[hbt]
\centering
\begin{tabular}{||c|c|c|c||}
\hline\hline
 & \multicolumn{3}{c|}{$\sigma$ (${\mathrm{MeV}}^{-1}$) }\\
 \hline
$\sqrt{-t}$ (MeV) & 1st Min & 2nd Min & 3rd Min\\
\hline
$100$ & 13.5 & & \\
$141$ & 10.5 & 21.0 & \\
$223$ & 8.5 & 17.0  & 25.5\\
$264$ & 8.0 & 16.0  & 24.0\\
$316$ & 7.5 & 15.0  & 22.5\\
$707$ & 7.0 & 13.5  & 20.0\\
\hline\hline
\end{tabular}
\caption{ Positions of minima in the diffraction pattern
for different $-t$ for the simulated hadron model DVCS amplitudes.}
\label{tab1}
\end{table}
From  Table \ref{tab1}, we can see that
for a fixed $(-t)$, higher minima appear at the positions integral
multiple of lowest minimum. This is  consistent with the single slit
diffraction law
for $n^{th}$ minima: $\sin \theta_n =n \lambda/w$ where $\lambda$ is the
wavelength of the light and $w$ is the slit width.
 Here  $\sigma$ plays the role of $\sin \theta_n$
(for large separation($L$) between the slit and the detector in a single
slit experiment, one can write
$\sin\theta_n \approx {\sigma_n\over L}$) and the ratio $\lambda/w$ is
determined by the value of $-t$.
 Positions of the minima do not depend on the helicity. The minima
 appear at the
 same places  for both helicity flip and non-flip processes.

We also observe a relation between the invariant momentum transfer
squared $-t$ and the distribution in $\sigma$:
%We introduce a parameter $L$, which is analogous to
%the distance of the detector from the slit in a single slit experiment, then
the first minimum in a diffraction pattern is determined by \be
\sin\theta_1={\sigma_1\over \sqrt{L^2+\sigma_1^2}}={\lambda\over w}
\ee 
where $\sigma_1$ is the position of the first minimum measured
from the center of the diffraction pattern. Introducing another
parameter $t_0$, with $-t >-t_0$ we write the right hand side i.e.,
the ratio $ {\lambda\over w}$ as, ${1\over \log(-t+t_0)}$ where we
have chosen $-t_0=2\times 10^4$ ${\mathrm{MeV}}^2$.
Once $t_0$ is fixed, the other
parameter $L$ can be found;  $\sigma_1=8.5$ ${\mathrm{MeV}}^{-1}$ for 
$-t=5\times 10^4$ $\mathrm{MeV}^2$ gives
\be 
{8.5\over \sqrt{L^2+8.5^2}}={1\over \log(3\times
10^4)}=0.2234
\nonumber 
\ee 
which gives $L^2\approx 1376$ ${\mathrm{MeV}}^{-2}$. Using this
value of $L$ we can compare with the other data:
\tablinesep=0.1in
\arraylinesep=0.1in
\extrarulesep=0.1in
\begin{table}[hbt]
\centering
\begin{tabular}{||c|c|c|c||}
\hline\hline
%$-t$ & $\sigma_1$ & ${\sigma_1\over \sqrt{L^2+\sigma_1^2}}$ &
%${1\over \log\sqrt{(-t+t_0)}}$\\
%\hline
%$7 \times 10^4$ & 8.0 & 0.425  & 0.426\\
%$1 \times 10^5$ & 7.5 & 0.403  & 0.408\\
%$5 \times 10^5$ & 7.0 & 0.380  & 0.352\\
%\hline\hline
$\sqrt{-t}$ & $\sigma_1$ & ${\sigma_1\over \sqrt{L^2+\sigma_1^2}}$ &
${1\over \log(-t+t_0)}$\\
\hline
$264$ & 8.0 & 0.211  & 0.213\\
$316$ & 7.5 & 0.198  & 0.204\\
$707$ & 7.0 & 0.185  & 0.176\\
\hline\hline
\end{tabular}
\caption{Comparison of our proposed formula with the data. $\sqrt{-t}$ 
is in MeV and the lengths are in ${\mathrm{MeV}}^{-1}$. }
\label{tab2}
\end{table}

Table \ref{tab2} shows  that our parameterization of ${\lambda\over
w}$ in terms of $-t$  is quite accurate.

If one Fourier transforms the change in transverse momentum
$\Delta_\perp$ to impact space $b_\perp$ \cite{bur1,bur2}, then one
would have the analog of a three-dimensional scattering center. In
this sense, studying the FT of the DVCS is very much like studying
the Lorentz-invariant optics of the proton. In our analysis we have
computed DVCS on an electron at $O(\alpha)$ and its FT. Thus we have
obtained the diffractive optics of the quantum fluctuations of an
electron.

It is interesting to recall that the scattering amplitude
corresponding to an absorptive (i.e., negative imaginary) potential
which is confined to a sphere of finite radius exhibits a diffraction
pattern. For a classic treatment, see R. J. Glauber's lectures
\cite{glauber}. For another example of diffraction patterns in the
angular distribution of elastic proton-nucleus scattering
%at low energies
using
%an optical potential in coordinate space,
a multiple-scattering approach, see Ref. \cite{optical}. It is
worthwhile to remember that in these examples the optical potential
is complex. In our case, we  observe diffraction patterns when we
perform Fourier Transforms of real functions.

%%%%%%%%%%%%%%%%%%%%%%%%%%%%%%%%%%%%%%%%%%%%%%%%%%%%%%%%%%%%%%%%%%%
\section{LFWF for meson in Holographic QCD and the DVCS amplitude}
%%%%%%%%%%%%%%%%%%%%%%%%%%%%%%%%%%%%%%%%%%%%%%%%%%%%%%%%%%%%%%%%%%%
\begin{figure}[t]
\centering
\psfrag{F1}{$\chi$}
\psfrag{S}{$\sigma$}
\psfrag{B}{$|b_\perp|$}
\includegraphics[width=7cm,height=7cm,clip]{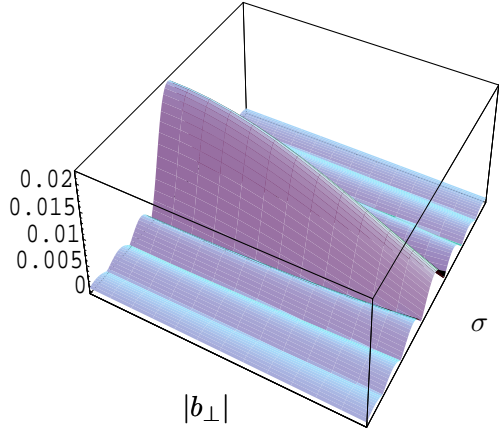}
%\hspace{0.2cm}%
%\includegraphics[width=7cm,height=7cm,clip]{adswf_v2.eps}
\caption{\label{fig11} (Color online) The ground state ( $L=0 ,k=1$) of  two parton
holographic light front wave function in 3D space. We have taken
$\Lambda_{QCD}=0.32 $ GeV. Here $|b_\perp|$ runs from 0.001 to 6.0 
$\mathrm{GeV}^{-1}$ and $\sigma$ from -25 to 25.}
\end{figure}

The normalized  holographic QCD LFWF for the meson  ($ q {\bar q}$)
from
%AdS/QCD
AdS/CFT derived in ref. \cite{tera} is \be
\Psi_{L,k}(x,b_\perp)=B_{L,k}\sqrt{x(1-x)}J_L(\xi
\beta_{L,k}\Lambda_{QCD}) \ee where
$B_{L,k}=\Lambda_{QCD}\big[(-1)^L \pi J_{1+L}
(\beta_{L,k})J_{1-L}(\beta_{L,k})\big]^{-1/2}$, $\xi=\sqrt{{x (1-x}}
|b_\perp|$, and $\beta_{L,k}$ is the $k$-th zero of Bessel function
$J_L$. For ground state $L=0,k=1$ and we have \be \phi(x,b_\perp)=
\Psi_{0,1}(x,b_\perp)=\Lambda_{QCD}\sqrt{x(1-x)}
{J_0(\xi\beta_{0,1}\Lambda_{QCD}) \over \sqrt{\pi}
J_1(\beta_{0,1})}\label{adswf} \ee The corresponding momentum space
LFWF is \cite{tera} \be \psi(x, \kappa_\perp) = \sqrt{4\pi^2}\int
d^2 {b_{1}}_\perp ~e ^{-i {b_{1}}_\perp \cdot \kappa_\perp} ~
\phi(x, {b_{1}}_\perp)~.\label{psi} \ee

In Fig. \ref{fig11}, we have plotted the two-parton bound state
holographic LFWF from the AdS/CFT correspondence  in 3D coordinate
space after taking the FT of Eq. (\ref{adswf}) with respect to $x$.
The ADS/CFT correspondence gives only the wavefunction for the
$q{\bar q}$ sector.  So, when we consider the DVCS amplitude with
this wavefunction we can have contribution only from the $2\to 2$
process. Then the DVCS amplitude can be written as \be M(
\Delta_\perp,\zeta) &=& -e_q^2~ \sum_{\sigma_1,\sigma_2}~
\int_\zeta^1 dx \left [ \frac{1}{x-\zeta+ i \epsilon} + \frac{1}{x -
i \epsilon} \right ] \int
d^2\kappa_\perp \nonumber \\
&~&\bigg [ \psi^*_{\sigma_1,\sigma_2}\left(\frac{x - \zeta}{1-\zeta},
    \kappa_\perp - \frac{1-x}{1-\zeta}\Delta_\perp \right)~ \psi_{\sigma_1,
      \sigma_2}(x,\kappa_\perp) ~~+ \nonumber \\
&~&  \psi^*_{\sigma_1,\sigma_2}\left(\frac{1-x}{1-\zeta},
    \kappa_\perp + \frac{1-x}{1-\zeta}\Delta_\perp \right)~ \psi_{\sigma_1,
      \sigma_2}(1-x, \kappa_\perp)~ \bigg ] ~.\label{Mq}
\ee

The transverse Fourier Transform of the DVCS amplitude gives the DVCS
amplitude in the transverse impact parameter space $b_\perp$.
\be
{\tilde A}(b_\perp, \zeta) = {1\over (2 \pi)^2}
\int d^2 \Delta_\perp ~ e^{i b_\perp \cdot
\Delta_\perp} ~ M (\Delta_\perp, \zeta)
%&=& \int dx ~F(x,\zeta)~ \int d^2\kappa^\perp ~ \int d^2 \Delta_\perp ~ e^{i
%b_\perp \cdot
%\Delta_\perp}~ \psi^*\left(\frac{x - \zeta}{1-\zeta},
%    \kappa_\perp - \frac{1-x}{1-\zeta}\Delta_\perp \right)~
%\psi(x,\kappa_\perp)~. \nonumber \\
\ee

\begin{figure}
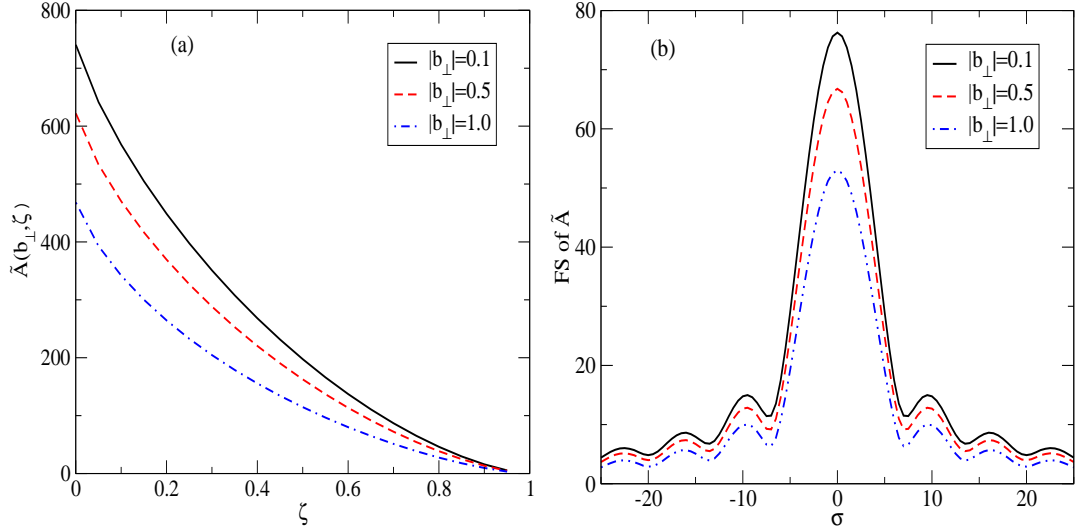

\includegraphics[width=7cm,height=7cm,clip]{longnew_fig16a.eps}%
\hspace{0.2cm}%
\includegraphics[width=7cm,height=7cm,clip]{longnew_fig16b.eps}
\caption{\label{adsdvcs} (Color online) (a) The DVCS amplitude vs. $\zeta$ and (b) the
 Fourier spectrum
of the DVCS amplitude in the $\sigma$ space
using the
light front wave function for meson  obtained from holographic QCD \cite{tera}.
 We have taken $\Lambda_{QCD}=0.32 $
GeV. Plots are in unit of $e_q^2$. $b^\perp$ is given in 
units of $\mathrm{GeV}^{-1}$. }
\end{figure}

Then
\be {\tilde A}(b_\perp, \zeta) &=&(2\pi)^4 \int
dx ~F(x,\zeta)~\frac{1-\zeta}{1-x}~
\big [
\phi^*(\frac{x-\zeta}{1-\zeta}, \frac{1-\zeta}{1-x} b_\perp) ~ \phi(x,
\frac{1-\zeta}{1-x}
b_\perp) \nonumber \\
&~&~~~~+\phi^*(\frac{1-x}{1-\zeta},-\frac{1-\zeta}{1-x} b_\perp)
\phi(1-x,-\frac{1-\zeta}{1-x} b_\perp)\big ]~.\label{Meqn}
\ee
%Thus in the transverse plane we reach a situation resembling the
% non-relativistic charge density.
where $F(x,\zeta)=-e_q^2 ({1\over x}+{1\over x-\zeta})$. Taking the
FT of this DVCS amplitude with respect to $\zeta$ we obtain the
amplitude in the 3-dimensional impact parameter space $\sigma,
b_\perp$. Substituting  the wavefunctions given in Eq.
(\ref{adswf}), we obtain \be A(\sigma,b_\perp)&=&{1\over 2\pi}\int
d\zeta e^{i\sigma\zeta}{\tilde A}(b_\perp,
\zeta)\nonumber\\
&=& 2 (2\pi)^4{\Lambda^2_{QCD}\over 2 \pi^2 J_1(\beta_{0,1})^2}\int_0^1 d\zeta e^{i\sigma\zeta}
\int_\zeta^1 dx ~F(x,\zeta)\sqrt{x(x-\zeta)}\big[J_0(X_1) J_0(X_2)\big]
\ee
where
\be
X_1&=&\sqrt{x ( 1-x)} \frac{1-\zeta}{1-x} |b_\perp |
~\beta_{0,1}\Lambda_{QCD}~,
\nonumber\\
X_2&=&\sqrt{(1-x)(x-\zeta)}  \frac{1}{1-x} | b_\perp |
~\beta_{0,1}\Lambda_{QCD}~.
\nonumber
\ee
In Fig. \ref{adsdvcs} we show the FS of the DVCS amplitude in $\sigma$ space
for different fixed values of $| b_\perp |$. Again we see the diffraction
pattern in the $\sigma$ space.

%%%%%%%%%%%%%%%%%%%%%%%%%%%%%%%%%%%%%%%%%%%%%%%%%%%%%%%%%%%%%%%%%%%%%%%%%%%
\section{Summary and Conclusions}
%%%%%%%%%%%%%%%%%%%%%%%%%%%%%%%%%%%%%%%%%%%%%%%%%%%%%%%%%%%%%%%%%%%%%%%%%%%
The deeply virtual Compton scattering process $\gamma^* p \to \gamma
p$ provides a direct window into hadron substructure which goes well
beyond inclusive measurements. The DVCS amplitude factorizes into  the
convolution of a hard perturbative amplitude, corresponding to
Compton scattering on a quark current, with the initial and final
state light-front wavefunctions of the target hadron. The LFWFs
provide a general frame-independent representation of  relativistic
composite hadrons, and they are universal and process independent.

In this paper, we have shown that the Fourier transform of the DVCS
amplitude with respect to the skewness variable $\zeta$ gives
information of the proton structure in longitudinal impact parameter
space $\sigma = {1\over 2}P^+ b^-$.

As an illustration of our the general framework, we have worked with
a simple relativistic spin $1/2$ system, namely the quantum
fluctuations of a lepton at one loop in QED. The different Fock
components of the LFWFs in this case can be obtained from
perturbation theory. Our calculation is exact to $O(\alpha)$. This
one-loop model provides a transparent basis for understanding the
structure of more general bound-state systems. By differentiating
the wavefunctions for the electron with respect to the square of the
bound state mass $M^2$, we have simulated bound state valence
wavefunctions.

We have noted that there are two different types of overlaps
contributing to DVCS when $\zeta$ is nonzero, namely a parton number
conserving diagonal $2 \rightarrow 2$ overlap and a $3 \rightarrow
1$ overlap when an electron-positron pair of the initial state is
annihilated. In fact, both these contributions are necessary in
order to obtain $\zeta$ independent form factors by taking the $x$
moment of the GPDs. This invariance is due to the Lorentz frame
independence of the light-front Fock representations of space-like
local operator matrix elements and it reflects the underlying
connections of Fock states with different parton numbers implied by
the equation of motion. The Fourier transform of the amplitude with
respect to $\zeta$ involves both type of contributions in different
kinematical regions.

We have introduced the light-cone longitudinal distance $\sigma= P^+
b^-/2$ and have shown the $\sigma$ dependence of the real and
imaginary part of the DVCS amplitude. The DVCS amplitude in $\sigma$
space represents an interference of the initial and final state
LFWFs. We have also shown the $\sigma$ dependence of the LFWFs
themselves.

We have exhibited the light-front coordinate space structure of our
model wavefunctions by performing the Fourier transform in the
longitudinal and transverse momentum space. The wavefunctions
exhibit diffraction patterns in the longitudinal coordinate space.
We have presented the FS of the real part of the DVCS amplitude as
well as the structure function in the models. The corresponding
imaginary part of the model DVCS amplitudes vanish.

Very recently, valence parton bound state holographic LFWFs from the
AdS/CFT correspondence  have been given \cite{tera}. We have
presented these wavefunctions for the meson in full three
dimensional light-front coordinate space. We have also calculated
the real part of the  DVCS amplitude in the holographic model in
light front longitudinal space for specific choices of the impact
parameter ($b_\perp$). Again one observes diffraction patterns. Note
that the imaginary part of the DVCS amplitudes vanish also in this
model.

Our analysis is the first to examine the longitudinal light cone
coordinate $\sigma = b^-P^+/2$ dependence of LFWFs and DVCS
amplitudes. Our results for the DVCS amplitude in $\sigma$ are
analogous to diffractive scattering of a wave in optics where the
$\sigma$ distribution senses  the size of the one-dimensional
scattering center.  Thus studying DVCS $\gamma^* p \to \gamma p$ in
light-front longitudinal coordinate space is very much like studying
the Lorentz-invariant optics of the proton.
%%%%%%%%%%%%%%%%%%%%%%%%%%%%%%%%%%%%%%%%%%%%%%%%%%%%%%%%%%%%%%%%%%%%%%%%
\section{acknowledgments}
%%%%%%%%%%%%%%%%%%%%%%%%%%%%%%%%%%%%%%%%%%%%%%%%%%%%%%%%%%%%%%%%%%%%%%%%%
This work was supported, in part, by the US Department of Energy
grant Nos. DE-AC02-76SF00515 and DE-FG02-87ER40371. This work was
performed in part, under the auspices of the US Department of Energy
by the University of California, Lawrence Livermore National
Laboratory under contract No. W-7405-Eng-48. This work was also
supported in part by the Indo-US Collaboration project jointly
funded by the U.S. National Science Foundation (NSF) (INT0137066)
and the Department of Science and Technology, India (DST/INT/US
(NSF-RP075)/2001). The work of DC was supported in part by US
Department of Energy under grant No. DE-FG02-97ER-41029. We thank
Professor Guy de Teramond for helpful conversations.
%%%%%%%%%%%%%%%%%%%%%%%%%%%%%%%%%%%%%%%%%%%%%%%%%%%%%%%%%%%%%%%%%%%%%%%%%
\appendix
%%%%%%%%%%%%%%%%%%%%%%%%%%%%%%%%%%%%%%%%%%%%%%%%%%%%%%%%%%%%%%%%%%%%%%
\section{Relation between Burkardt and Soper densities}\label{density}
%%%%%%%%%%%%%%%%%%%%%%%%%%%%%%%%%%%%%%%%%%%%%%%%%%%%%%%%%%%%%%%%%%%%%%
The off-forward parton distribution appropriate for a spin-zero
meson in the valence approximation is defined by \be H(x,\zeta,t)&=&
\int {dy^-\over 8\pi} e^{i xP^+y^-/2}\langle
P^\prime\mid{\bar\psi}(0)\gamma^+\psi(y^-)\mid P\rangle_{y^+=0}\nonumber \\
&=& \int_0^1 dz~ \delta(x-z)\theta(z-\zeta)\int
d^2k_\perp \psi^*({z-\zeta\over 1-\zeta},k_\perp -
{1-z\over 1-\zeta}\Delta_\perp)\psi(z,k_\perp)~.
\ee
For skewness $\zeta=0$,
\be
H(x,\zeta=0,t)=H(x,\Delta_\perp) &=&\int d^2k_\perp
\psi^*(x, k_\perp -(1-x)\Delta_\perp)
\psi(x,k_\perp) \nonumber\\
&=& \int {d^2b_\perp \over (2\pi)^2} e^{-i(1-x)b_\perp\cdot
\Delta_\perp} \phi^*(x,b_\perp)\phi(x,b_\perp) \ee where the FT of
the wavefunction  is defined as \be \psi(x,k_\perp)= \int
{d^2b_\perp \over (2\pi)^2} e^{-ib_\perp\cdot k_\perp}
\phi(x,b_\perp)~. \ee The transverse Fourier transform of the zero
skewness off-forward parton distribution $H$ \cite{bur1} yields the
impact parameter density function \be \int {d^2 \Delta_\perp\over (2
\pi)^2} e^{i\eta_\perp\cdot \Delta_\perp} H(x,\Delta_\perp) &=& \int
d^2b_\perp \delta^2[(1-x)b_\perp-\eta_\perp]\phi^*(x,b_\perp)
\phi(x,b_\perp) \nonumber\\
&=& {\rho(x,{\eta_\perp\over 1-x})\over 1-x}, \ee where
$\rho(x,b_\perp)$ is the Soper density defined in Eq. (5) of Ref.
\cite{soper}. We thus find that the density obtained by Burkardt is
the same as the Soper density.
%%%%%%%%%%%%%%%%%%%%%%%%%%%%%%%%%%%%%%%%%%%%%%%%%%%%%%%%%%%%%%%%%%%%
\section{Regulators}\label{cutoff}
%%%%%%%%%%%%%%%%%%%%%%%%%%%%%%%%%%%%%%%%%%%%%%%%%%%%%%%%%%%%%%%
Let us consider the real part of the DVCS amplitude for electron in one-loop
QED given by
\be
Re ~ M(\zeta, \Delta_\perp)  &=& - e^2 \int_0^\zeta ~ dx ~
F^{31}(x,\zeta,\Delta_\perp) ~ \left [ \frac{1}{x} ~ + ~
 \frac{1}{x-\zeta} \right ] \nonumber \\
&~& - e^2 \int_\zeta^1 ~dx~ F^{22}(x,\zeta, \Delta_\perp)~ \left [
\frac{1}{x}  ~+~  \frac{1}{x - \zeta} \right ]~
\ee
which results after performing the transverse momentum integration. As
described in the text, we use an ultraviolet  cutoff $\Lambda$ on the
transverse momentum.

The integrands for various DVCS amplitudes may exhibit singular
behavior when  $x$ is near the end points. We also have a potential
singularity when $ x \rightarrow \zeta$ which can be regulated by
using the principal value prescription. In the numerical work we
implement the principal value prescription by employing suitable
regulators and ensuring regulator independence in the limit where
the regulator vanishes. We also note that we eventually integrate
over $\zeta$ which ranges between $\zeta_{min},$ which is close to
zero and $\zeta_{max},$ which is determined by $-t$ and approaches
$1$ in the limit $-t \rightarrow \infty $.

The light-cone momentum fractions must remain positive. Let
$\zeta_{min} = \zeta + \epsilon/2$ for the second integral.  Since
$\zeta_{max} = 1 - \epsilon$, to make sure that $x$ remains greater
than $\zeta$ in the second integral, we choose $x_{max}$ in the
second integral to be $ 1 - \epsilon/2$.
Thus the regulated integral is
\be
Re ~ M(\zeta, \Delta_\perp)  &=& - e^2 \int_{\epsilon/2}^{
\zeta - \epsilon/2}~ dx ~
F^{31}(x,\zeta,\Delta_\perp) ~ \left [ \frac{1}{x} ~ + ~
 \frac{1}{x-\zeta} \right ] \nonumber \\
&~& - e^2 \int_{\zeta+ \epsilon/2}^{1 - \epsilon/2} ~dx~ F^{22}(x,
\zeta, \Delta_\perp)~ \left [
\frac{1}{x}  ~+~  \frac{1}{x - \zeta} \right ]~.   \label{cut}
\ee

It is important to note that when $\zeta$ is small, significant
contribution to the integral comes from the second term that
involves $F^{22}$ and when $\zeta$ is large, significant
contribution to the integral comes from $F^{31}$. For the
helicity-flip case, both $F^{22}$ and $F^{31}$ vanish as $x
\rightarrow 0$. For the helicity non-flip amplitude, $F^{31}$
vanishes as $x \rightarrow 0$, but $F^{22}$ is finite as $x
\rightarrow 0$. Thus the only potential problem at small $x$ occurs
in Eq. (\ref{cut}) for the {\em second term} involving $F^{22}$ when
$\zeta$ is small and we obtain a logarithmic divergence due to this
problem. This is directly related to the non-vanishing of the
electron wave function as $ x \rightarrow 0$. For helicity non-flip,
the function $F^{22}$ also diverges as $x \rightarrow 1$. In this
region, the DVCS amplitude also receives contributions from the
single-particle sector of the Fock space which we do not take into
account in the present calculation. If one uses an invariant mass
cutoff, the divergences at $x=0$ and $x=1$ would have been regulated
by non-zero electron and photon masses respectively. These
regulators are not mandatory in our present calculations, and we
have employed the  simpler regulators as described above.
%%%%%%%%%%%%%%%%%%%%%%%%%%%%%%%%%%%%%%%%%%%%%%%%%%
\section{An illustrative model}\label{step}
%%%%%%%%%%%%%%%%%%%%%%%%%%%%%%%%%%%%%%%%%%%%%%%%%
Let us start  from the expression for the Fourier transform
\be
A(\sigma)={1\over 2 \pi}\int d\zeta~e^{i\sigma\zeta} M(\zeta)~.
\ee
The function $A(\sigma)$ is in coordinate space and the function
$M(\zeta)$ is in momentum space.

Let us approximate $M$ by the following function:
%Thus our step function is in momentum space.

\be
M(\zeta)&=& M_0~~~ {\rm for}~~ 0 < \zeta < \zeta_{max}\nonumber \\
 &=& 0~~~ {\rm for} ~~ \zeta > \zeta_{max}
\ee

One can use this step function to approximate a DVCS amplitude in
which the dependence of the DVCS amplitude in $\zeta$ is almost
flat; in such a case we can take  $M_0  = [M(\zeta=0) +
M(\zeta=\zeta_{max}]/2$.

We have,
\be
A(\sigma)&=&{M_0\over 2 \pi}\int_0^{\zeta_{max}} d\zeta
~e^{i\sigma\zeta}\nonumber\\
&=&{M_0\zeta_{max}\over 2 \pi}~
{\sin(\sigma\zeta_{max}/2)\over \sigma\zeta_{max}/2} ~e^{i\sigma\zeta_{max}/2}.
\ee
In this case, note that the cosine and sine transforms are completely in
phase and the phase of the Fourier transform does not contain any extra
information. The amplitude (i.e., the Fourier Spectrum) is given by
\be
\mid A (\sigma) \mid ~= ~ {M_0\zeta_{max}\over 2 \pi}~
 { \mid \sin(\sigma\zeta_{max}/2) \mid \over \sigma\zeta_{max}/2} ~.
\ee

The magnitude of the peak of the diffraction pattern
\be
A(\sigma)_{max}={M_0\zeta_{max}\over 2\pi}
\ee
and the first diffraction minimum occurs at
\be
\sigma_1={2\pi \over \zeta_{max}}~.
\ee

In the case of the DVCS amplitude whose functional dependence on
$\zeta$ is very weak, we can further predict the position of the
minima as follows. The extension of the function $\zeta_{max}$ is
given by \be \zeta_{max}={-t\over 2 M^2}\Big (\sqrt{1+{4M^2\over
-t}}-1\Big )~. \ee Thus we find a precise relation between the
minima of the diffraction pattern and $-t$. Since $\sigma_1$ is
inversely proportional to $\zeta_{max}$ which in turn increases with
$-t$, the inward movement of the first minimum with increasing $-t$
is readily explained. On the other hand the peak height is a
 product of $M_0$ and $\zeta_{max}$. The amplitude $M_0$ decreases
 %monotonously
 monotonically
 with increasing $-t$. On the other hand $\zeta_{max}$ increases
 with increasing $-t$. Thus the peak height has
 %non-monotonous
 non-monotonic
 behavior with
 respect to $-t$.

In Table \ref{step1} we compare  the numbers with the case of the
real part of the helicity non-flip amplitude presented in Figs.
\ref{fig10}(a) and (b). Similarly one can also obtain
%approximate
estimates for the helicity flip amplitudes. \tablinesep=0.1in
\arraylinesep=0.1in \extrarulesep=0.1in
\begin{table}[hbt]
\centering
\begin{tabular}{||c|c|c|c|c||}
\hline
$\sqrt{-t}$ &$\zeta_{max}$& $M_0$ & Peak (${M_0\zeta_{max}\over 2 \pi}$)& First Minimum ($\sigma_1={2\pi\over \zeta_{max}}$) \\
\hline
100 & 0.48 & 0.04 & $3\times 10^{-3}$ & 13.09\\
316 & 0.84 & 0.03  & $4\times 10^{-3}$ & 7.48\\
707 & 0.96 & 0.0175  & $2.7\times 10^{-3}$ & 6.54\\
\hline
\end{tabular}
\caption{Simplified  approximation for the real part of the helicity
non-flip DVCS amplitude shown in Figs. \ref{fig10} (a) and (b). The 
energy quantities are given in MeV and lengths in ${\mathrm{MeV}}^{-1}$.}
\label{step1}
\end{table}

The essential ingredients for the diffraction pattern in the Fourier
Spectrum are two
characteristics of the DVCS amplitudes in the variable $\zeta$:\\
$\bullet$ A step (i.e., a sharp rise) and\\
$\bullet$ a plateau.\\
These are the essential characteristics of a function which is almost a
constant that seems to be
shared by the DVCS amplitudes that produce a diffraction pattern in the FS.
The imaginary part of the helicity-flip  DVCS amplitude for the electron
 state (Fig. \ref{fig3}(a)) lacks these properties
and we do not observe any diffraction pattern in the
corresponding Fourier Spectrum (Fig. \ref{fig5}(a)).

It is interesting to note that the simple model we have discussed in this
appendix appears in antenna theory \cite{bracewell}. In the case of an
aperture for which a uniform electric field is maintained over a finite
distance, outside of which the field is zero, the angular spectrum which is
the Fourier Spectrum of the aperture field distribution exhibits
the diffraction pattern discussed in this appendix.
%%%%%%%%%%%%%%%%%%%%%%%%%%%%%%%%%%%%%%%%%%%%%%%%%%%%%%%%%%%%%%%%%%%%%
\section{DVCS amplitude in three dimensions}
%%%%%%%%%%%%%%%%%%%%%%%%%%%%%%%%%%%%%%%%%%%%%%%%%%%%%%%%%%%%%%%%%%%%%%%%%%

The significance of the amplitude in the boost-invariant
$\sigma$ space can also be explained in the following way. The Dirac and
Pauli form factors $F_1(t)$ and $F_2(t)$ respectively,
can be expressed in terms of the helicity non-flip part of the off-forward
matrix element,

Let us consider the dressed electron in the frame $\zeta=0$.
The form factor can be written as \cite{soper}
\be
F(t)=\int_0^1 dx \int d^2 b^\perp e^{-i \Delta^\perp
\cdot b^\perp} {\mid
\tilde \psi_2(x,b^\perp) \mid }^2
\ee
The LFWFs in the mixed representation $x,b^\perp$ are given by Eq.
(1) of \cite{tera}. Note that the LFWFs are zero outside
the region $0<x<1$. We denote :
\be
\Psi_2(x, b^\perp)= \tilde \psi_2(x,b^\perp) \theta(x) \theta(1-x)\nonumber\\
\Psi_3(x_1,x_2,b^\perp_1,b^\perp_2)= \tilde \psi_3(x_1, x_2, b^\perp_1,
 b^\perp_2) \theta(x_1) \theta(x_2) \theta(1-x_1) \theta(1-x_2)
\ee
for the two particle LFWF and similarly $\Psi_2$ for the three-particle
wave function.
We take FT of the LFWFs $\Psi$ with respect to $x$ ; and define
\be
\Phi_n(\sigma_i, b^\perp_i)= \{\Pi_{i=1}^{n-1}\int_{-\infty}^{+\infty}
d \sigma_i e^{-i \sigma_i x_i} \}
\Psi_n(x_i,b^\perp_i)
\ee
where $\sigma_i$ are the boost invariant longitudinal distance on the light
cone, conjugate to $x_i={k_i^+/P^+}$. There are $n-1$ independent
$\sigma_i$ as well as $b^\perp_i$. In terms of these, we can write,
\be
F(t)&=&\int dx \int d^2 b^\perp e^{-i \Delta^\perp
\cdot b^\perp} \int d \sigma_1 \int d \sigma_2 e^{i {\sigma}_1 x} e^{-i {\sigma}_2 x}
\Phi_2^*(\sigma_1,b^\perp) \Phi_2 (\sigma_2, b^\perp)\nonumber\\
&=&
2 \pi \int d^2 b^\perp e^{-i \Delta^\perp
\cdot b^\perp} \int d \sigma {\mid
\Phi_2(\sigma,b^\perp) \mid }^2
\ee
Note that as $\Phi_n$ are the FT of the wave functions $\Psi_n$ rather
than $\tilde \psi_n$, it is mathematically correct to take the $x$-
integrals from $-\infty$ to $+ \infty$.
When $\zeta$ is non-zero, the form factor receives contributions from
$2-2$ and $3-1$ components of the GPDs $H$ and $E$. They can be obtained from,
\be
\int &dx &[ F^{22}_{++}(x,\zeta,t) \theta(x-\zeta) +F^{31}_{++} (x, \zeta,t)
\theta(\zeta-x)] \nonumber\\&&~~~~~~~~~~=
\sqrt{1-\zeta} F_1(t)-{\zeta^2\over 4 \sqrt{1-\zeta}} F_2 (t)\nonumber\\
&\approx&\int_0^1 dx \int d^2 b^\perp e^{-i \Delta^\perp \cdot b^\perp}
\Big [ \sqrt{1-\zeta} \tilde \psi_3^{\uparrow}(x,1-\zeta,\zeta-x,-b^\perp)
\theta(\zeta-x)\nonumber\\&&~~~~~~~~~+\tilde \psi_2^{*,\uparrow}(x',
{b^\perp\over 1-x'})\tilde \psi_2^{\uparrow} (x,{b^\perp\over 1-x'})
{(1-\zeta)^2 \over (1-x)^2} \theta(x-\zeta)\Big ].
\label{snf}
\ee
\be
\int_0^1 &dx&[ F^{22}_{+-}(x, \zeta,t)\theta(x-\zeta) +F^{31}_{+-} (x,
\zeta,t) \theta(\zeta-x)] \nonumber\\&&~~~~~~~~~~~~~~~=
 {1\over \sqrt{1-\zeta}} {(\Delta^1-i \Delta^2) (1-{\zeta/2})
\over 2 M} F_2 (t)\nonumber\\
&\approx&\int_0^1 dx \int d^2 b^\perp e^{-i \Delta^\perp \cdot b^\perp}
\Big [ \sqrt{1-\zeta} \tilde \psi_3^{\downarrow}(x,1-\zeta,\zeta-x,-b^\perp)
\theta(\zeta-x)\nonumber\\&&+\tilde \psi_2^{*,\uparrow}(x',{b^\perp\over 1-x'})
\tilde \psi_2^{\downarrow} (x,{b^\perp\over 1-x'}) {(1-\zeta)^2
\over (1-x)^2} \theta(x-\zeta)\Big ].
\label{sf}
\ee
$x'={x-\zeta\over 1-\zeta}$. Here we have suppressed the explicit helicity
indices and used
\be
F_1(t)=\int_0^1 {H(x, \zeta, t)\over 1-{\zeta\over 2}}, \nonumber\\
F_2(t)=\int_0^1 {E(x, \zeta, t)\over 1-{\zeta\over 2}},
\ee
The form factors $F_1(t)$ and $F_2(t)$ can be obtained in terms of
overlaps of LFWFs in the mixed representation $\tilde \psi_n$ from Eq.
(\ref{snf}) and (\ref{sf}). Note that as the argument of the wave
functions are ${b^\perp\over 1-x'}$, these equations cannot be expressed
as an overlap of the FT wave functions $\Phi_n(\sigma_i, b^\perp_i)$
in position space.

However, in the mixed representation, one can write,
\be
F_1(t)&&= \int d^2 b^\perp e^{i \Delta^\perp \cdot b^\perp }
\Big [ \int_0^\zeta dx R_{31} (x, \zeta, b^\perp)+
\int_\zeta^1 dx R_{22} (x, \zeta, b^\perp) \Big ]\nonumber\\&&
=\int_0^1 dx \int d^2 b^\perp e^{i \Delta^\perp \cdot b^\perp }
R(x,\zeta, b^\perp)=
\int_0^1 dx  \int d^2 b'^\perp e^{i \Delta'^\perp \cdot
b'^\perp }\rho (x, b'^\perp);
\label{sop1}
\ee
\be
F_2(t)&& = \int d^2 b^\perp e^{i \Delta^\perp \cdot b^\perp }
\Big [ \int_0^\zeta dx \tilde R_{31} (x, \zeta, b^\perp)+
\int_\zeta^1 dx \tilde R_{22} (x, \zeta, b^\perp) \Big ]\nonumber\\&&
=\int_0^1 dx \int d^2 b^\perp e^{i \Delta^\perp \cdot b^\perp }  \tilde R(x,\zeta, b^\perp)=
\int_0^1 dx  \int d^2 b'^\perp e^{i \Delta'^\perp \cdot
b'^\perp }\tilde \rho(x, b'^\perp).
\label{sop2}
\ee
$R(x,\zeta,b^\perp)$ and  $\tilde R(x,\zeta,b^\perp) $ can be obtained
in terms of off-diagonal
overlaps of LFWFs $\tilde \psi_n(x_i,b^\perp_i) $ and can be obtained
from the above equations. $\rho(x,b'^\perp)$ and $\tilde \rho(x,b'^\perp)$
are Soper's distributions in the frame $\zeta=0$. It can be shown that
\be
\tilde \rho(x,b^\perp)=-2i M  {\partial \over \partial b}
{(b^1+i b^2) \over b}\rho(x,b^\perp).
\ee
Eqs. (\ref{sop1}) and (\ref{sop2}) show the relation between the generalized
correlation functions $R$ and $\tilde R $ with the Soper distribution
due to covariance of the form factor. However, the functions $R$ and
$\tilde R$ do not have a probability interpretation, unlike Soper's distribution.

In the imaginary part of the DVCS amplitude, we have the GPDs integrated
with a delta function,
\be
Im[M^{++}]= N \int_0^1 & dx & \Big [\delta(x-\zeta)
{\sqrt{1-\zeta} \over 1-{\zeta\over 2}}\ H_{(2\to
2)}(x,\zeta,t)~
\nonumber\\&&~~~-~ {\zeta^2 \over 4 (1-{\zeta\over 2}){\sqrt{1-\zeta}}}
E_{(2\to 2)}(x,\zeta,t) \delta(\zeta-x) \Big ]
\ee
In terms of the correlation functions defined above, this can be written as
\be
Im[M^{++}]&=& N\int d^2 b^\perp e^{i \Delta^\perp \cdot b^\perp}
\int_0^1 dx  \Big \{ \sqrt{1-\zeta} \Big [ \theta(\zeta-x) R_{31}
(x,\zeta,b^\perp) \nonumber\\&& +
\theta(x-\zeta) R_{22}(x,\zeta,b^\perp) \Big ] \delta(x-\zeta)
-{\zeta^2\over 4 \sqrt{1-\zeta}}
\Big [ \theta(\zeta-x) \tilde R_{31}(x,\zeta,b^\perp) \nonumber\\&& +
\theta(x-\zeta) \tilde R_{22}(x,\zeta,b^\perp) \Big ] \delta(x-\zeta)
\Big \} \nonumber\\
&=&\int d^2 b^\perp e^{i \Delta^\perp \cdot b^\perp}\int_0^1 dx \Big
\{ \sqrt{1-\zeta} R (x,\zeta,b^\perp)  \delta(x-\zeta)
\nonumber\\&&~~~~~~~~~-{\zeta^2\over 4 \sqrt{1-\zeta}} \tilde R
(x,\zeta,b^\perp) \delta(x-\zeta) \Big \} \ee $N$ is the
normalization constant. Integrating over $x$ we obtain \be
Im[M^{++}]&=& N\int d^2 b^\perp e^{i \Delta^\perp \cdot b^\perp}
\Big \{ \sqrt{1-\zeta} R (x=\zeta,b^\perp)
\nonumber\\&&~~~~~~~~~-{\zeta^2\over 4 \sqrt{1-\zeta}} \tilde R
(x=\zeta,b^\perp) \Big \} \ee Thus, the FT of the imaginary part of
the DVCS amplitude with respect to $\Delta^\perp$ gives both
$R(x,\zeta, b_\perp)$ and  $\tilde R(x, \zeta,b_\perp)$ where $x$ of
the struck parton is now fixed at $x=\zeta$. These, in turn, are
related to Soper's distributions $\rho(x,b^\perp)$ through Eq.
(\ref{sop1}) and (\ref{sop2}). This is a mixed coordinate and
momentum space representation. The above relation can be generalized
to a hadron in a model independent way. Introducing the complete 3D
spatial amplitude ${\bar \rho} (\sigma, b^\perp)$ at fixed light
front time $\tau$,  we can write, \be Im[M^{++}]&= & N \int d \sigma
e^{-i \sigma \zeta} \int d^2 b^\perp e^{-i \Delta^\perp \cdot
b^\perp}~ {\bar \rho} (\sigma, b^\perp). \ee Here $\sigma$ is
conjugate to $\zeta$. Note that as we are at fixed $\tau$ rather
than at fixed time, there is no conceptual problem due to Lorentz
boosts.

The physics of the real part of the DVCS amplitude is more involved.
However, it is related to the imaginary part by a dispersion
relation in $x$. The real part can be expressed in terms of the
densities $\rho$ and $\tilde \rho$ as well, however, it contains a
principal value (PV) integral over $x$. We can call the result of
the PV integral $\gamma(\zeta, b^\perp)$. Again, after taking a FT
in $\zeta$ we obtain the amplitude in full 3D coordinate space.

%%%%%%%%%%%%%%%%%%%%%%%%%%%%%%%%%%%%%%%%%%%%%%%%%%%%%%%%%%%%%%%%%%%%%%%%

%%%%%%%%%%%%%%%%%%%%%%%%%%%%%%%%%%%%%%%%%%%%%%%%%%%%%%%%%%%%%%%%%%%%%%%%%%%%%
\end{document}